\documentclass[11pt,a4paper]{article}

\usepackage{float}
\usepackage{amsmath,epsfig,amssymb,subfigure,bm,dsfont}
\usepackage{mathrsfs}
\usepackage{cuted}
\usepackage{physics}
\usepackage{amsmath}
\usepackage{mathtools,amssymb,lipsum, nccmath}
\usepackage{comment}
\usepackage{array}
\def\a{\alpha}

\def\e{\epsilon}

\def\<{\langle}
\def\>{\rangle}
\def\oo{\infty}

\usepackage{verbatim}

\usepackage{fancyhdr}
\usepackage[english]{babel}
\usepackage[utf8x]{inputenc}
\usepackage[T1]{fontenc}

\usepackage[a4paper,top=3cm,bottom=3cm,left=3cm,right=3cm,marginparwidth=1.75cm]{geometry}

\usepackage{amsmath}
\usepackage{graphicx}
\usepackage{url}
\usepackage[colorinlistoftodos]{todonotes}
\usepackage[colorlinks=true, allcolors=blue]{hyperref}

\usepackage{authblk}

\numberwithin{equation}{subsection}

\title{\textbf{Numerical prescriptions of early-time divergences of the in-in formalism}}

\author[1]{Duc Huy Tran}
\author[1,2]{Yi Wang}
\author[3]{Juanyi Yang}
\author[1,2]{Yuhang Zhu}
\affil[1]{Department of Physics, The Hong Kong University of Science and Technology, Clear Water Bay, Kowloon, Hong Kong, P.R. China}
\affil[2]{Jockey Club Institute for Advanced Study, The Hong Kong University of Science and Technology, Clear Water Bay, Kowloon, Hong Kong, P.R. China}
\affil[3]{Department of Physics, The University of Hong Kong, Pok Fu Lam, Hong Kong, P.R. China}

\date{}

\begin{document}
\maketitle

\begin{abstract}
In quantum field theory, the in and out states can be related to the full Hamiltonian by the $i\epsilon$ prescription. A Wick rotation can further bring the correlation functions to Euclidean spacetime where the integrals are better defined. This setup is convenient for analytical calculations. However, for numerical calculations, an infinitesimal $\epsilon$ or a Wick rotation of numerical functions are difficult to implement. We propose two new numerical methods to solve this problem, namely an Integral Basis method based on linear regression and a Beta Regulator method based on  Cesàro/Riesz summation. Another class of partition-extrapolation methods previously used in electromagnetic engineering is also introduced. We benchmark these methods with existing methods using in-in formalism integrals, indicating advantages of these new methods over the existing methods in computation time and accuracy.
\end{abstract}

\section{Introduction} 
\label{sec:introduction} 

Cosmological correlation functions play a central role in understanding the dynamics and matter content of cosmological inflation. The in-in formalism can be applied to calculate these correlators \cite{schwinger1960special,bakshi1963expectation,BakshiPradipM1963EVFi,Weinberg:2005vy} (see \cite{Chen:2010xka, Wang:2013zva, zhongzhi2017} for reviews). The process of computing correlators for a given inflation model can be laborious. Unlike the correlator lives in the flat space-time following the Lorentz-covariance, there is no time-translation invariance during inflation. Consequently, time integrals with diverse integrands arise. 

Given the initial conditions of Bunch-Davies vacuum, mode functions oscillate rapidly at the early time (e.g.~$\sim e^{ik\tau}$), this leads us to calculate the highly oscillating integrals in the form of $ \int_{-\infty}^{\tau_{0}}g(\tau)e^{i\omega\tau}d\tau~$,
where $ g(\tau) $ is a non-oscillating or slowly oscillating function in general.
When more vertices are involved, the integration on the time domain becomes nested and more complicated. Each layer of integral can still be approximated in this form.
Upon obtaining the analytical expression of the mode functions, the standard procedure is to apply Wick rotation by rotating $\tau$ to the imaginary axis. After the Wick rotation, the oscillatory integrand will decay exponentially and converge (see, for example \cite{chen2010quasi}). However, in some inflation models with non-trivial features, for example, \cite{chen2007large,arroja2011large,adshead2013bispectrum}, the evolution of modes cannot be solved analytically. In these cases where the expression of the integrand is numerical, it can be technically challenging to evaluate the integral by Wick rotation. If we directly put on a cut-off to avoid evaluating the highly oscillated integral at early time, a $\mathcal{O}(1)$ spurious contribution will be mistakenly included \cite{chen2007large}. 

There are several known methods that can effectively handle with this dilemma. One is to directly introduce a small damping factor $e^{-\beta\tau}$ by hand, which is able to eliminate the oscillatory tails at the early time \cite{chen2007large}. The second solution is through integration by parts to speed up the convergence \cite{chen2008generation}. Another elegant technique is to use the so called Hölder summation to regulate the divergent oscillatory tails \cite{junaid2015geometrical}.

In this paper, we first propose some new methods that have its own advantages in solving the divergent oscillatory tails. The first one is similar to the Hölder summation, that based on Cesàro/Riesz summation. Another method is through choosing a basis of functions to separate the possible divergent part and then evaluate the integral semi-analytically. We will also review the partition-extrapolation (PE) methods \cite{pemethodsreview,pemethodsreview2}, a set of very efficient methods which has been invented in the last century and made great success in the electric engineering area. This method takes a different approach by exploiting the knowledge of the asymptotic behaviour of the integrand to dramatically accelerate the convergence. To compare the performance of different methods and show their own advantages, we apply different typical methods to two examples (one with the numerical integrand) to evaluate their convergence speed as well as the computation time consumption.

This paper is organized as follows, we first give a brief introduction about different known methods in Section~\ref{knownmethods}. 
 In Section~\ref{othermethods}, we will introduce several innovative methods including the integral basis method and highly efficient PE methods.
 In Section~\ref{comparison}, we will compare the performance of different methods through two applications and discuss their strength and weaknesses. We conclude in Section~\ref{conclusions}.

\section{Summary of known methods}
\label{knownmethods}

In this section, we review several known methods to numerically evaluate the highly oscillating integrals.

\subsection{Damping factor method}
\label{dampingfactor}
By manually adding a small damping factor $\beta$ into the integrand \cite{chen2007large} like,
\begin{equation}
\label{dampingintegrand}
\int_{-\infty}^{\tau_{0}}g(\tau)e^{i\omega\tau}\times e^{\beta\omega(\tau-\tau_{0})}d\tau~,
\end{equation}
\noindent which is similar to rotating the variable into the imaginary plane $ \tau\to\tau(1-i\beta) $, the integrand will quickly converge when $\tau$ approaches to the infinity. However, since the non-oscillatory factor $ g(\tau) $ is not rotated, \eqref{dampingintegrand} is only approximately equivalent to the expected result from the Wick rotation. In addition, one should choose the damping factor $\beta$ carefully under the trade-off between accuracy and efficiency, which is illustrated in Appendix \ref{DFconv}.

\subsection{Boundary regulator method}
The second useful method for speeding up the convergence at the far past is through integration by parts (IBP)~\cite{chen2008generation}. In general, the behavior of the non-oscillatory part $g(\tau)$ can be approximated by some power law functions $g(\tau)\sim \tau ^p$ when $\tau\rightarrow -\infty$. Nevertheless, when the power index $p>0$, the $g(\tau)$ function itself will suffer divergence at the early time which indeed will slow down the speed and decrease the accuracy of the numerical evaluation. By performing the integration by part $n$ times with $n>p$, we are able to effectively suppress this kind of divergence. More specifically, after implementing IBP enough times, the original target integral is transformed into the below form which involves computing the numerical derivatives \cite{chen2008generation},
\begin{align}
\label{integrationbyparts}
    &\int_{-\infty}^{\tau_{0}}g(\tau)e^{i\omega\tau}d\tau  \notag\\
    &=\sum_{m=1}^{n}(-1)^{m-1}\left(\dfrac{1}{i\omega}\right)^{m}&e^{i\omega\tau}\dfrac{d^{m-1}g(\tau)}{d\tau^{m-1}}\bigg|_{-\infty}^{\tau_{0}}
    +(-1)^{n}\left(\dfrac{1}{i\omega}\right)^{n}\int_{-\infty}^{\tau_{0}}&\dfrac{d^{n}g(\tau)}{d\tau^{n}}e^{i\omega\tau}d\tau~.
\end{align}
The boundary terms at $ \tau=-\infty $  vanish with the help of the $ i\epsilon $ prescription. The remaining integral is more convergent than the original one because the degree of divergence has been lowered by $ n $ through applying the $ n $-th order derivative. If the degree of divergence is lowered to a negative value, the integral would automatically converge at the early stage without the necessity of numerically applying the $ i\epsilon $ prescription because the new integrand would be suppressed by a power-law like function. The contribution from the highly oscillatory part during the early time becomes negligible and can therefore be safely deserted. The challenge of the boundary regulator method is the numerical evaluation of derivatives, which requires more precision in numerical mode functions and sometimes can introduce slowdowns or artifacts if not taken carefully.

\subsection{Hölder summation method}
The third possible way is based on the Hölder summation of integrals. The $ (H,\alpha) $ sum of an integral $ \int_{-\oo}^{\tau_{0}}f(\tau)d\tau $ is defined as \cite{hardydivseries},
\begin{align}
    \label{Holdersum}
    &\underset{(H,\alpha)}{\int_{-\oo}^{\tau_{0}}} f(\tau)d\tau 
    \notag\\
    &\equiv\lim_{\tau\to-\infty}\dfrac{1}{\tau-\tau_{0}}\int_{\tau_{0}}^{\tau}d\tau^{(\alpha)}\dfrac{1}{\tau^{(\alpha)}-\tau_{0}}\cdots\int_{\tau_{0}}^{\tau^{(3)}}d\tau^{(2)}\dfrac{1}{\tau^{(2)}-\tau_{0}}\int_{\tau_{0}}^{\tau^{(2)}}d\tau^{(1)}\int_{\tau_{0}}^{\tau^{(1)}}d\tau^{(0)}f(\tau^{(0)})~.  
\end{align}
Then the value of the integral $ \int_{-\oo}^{\tau_{0}}f(\tau)d\tau $ with the $ i\epsilon $ prescription is equivalent to its $ (H,\alpha) $ sum for a large enough non-negative integer $ \alpha $. Superficially, the original one-dimension integral has been transformed into the higher-dimensional one, with the tremendous increase in the numerical time complexity. Nevertheless, we noticed here is a numerical trick that can efficiently handle this problem. The idea is to convert integration into solving a corresponded ordinary differential equation that are able to reduce the evaluation time of the multi-dimensional integrals. We leave more details about this point in the Appendix.~\ref{ulftrick} and ~\ref{ulftrickwithode}~.

The Hölder method is firstly introduced in \cite{junaid2015geometrical}, where the authors used the $ (H,2) $ sum to calculate the target integrals. We provide a proof that, if the non-oscillatory factor $ g(\tau) $ has the form $ g(\tau)\sim\tau^{n} $ (which is often the case in the calculation of inflationary correlation functions), then $ (H,\alpha) $ is summable ($i.e.$ the limit in \eqref{Holdersum} converges) if and only if $ \alpha>n $. Technical proof is located at the Appendix \ref{H-C-Rconvergence}. In numerical calculation, given that the $ (H,\alpha) $ sum is convergent, we can safely choose a suitable early time cut-off $ \tau_{\text{early}} $ to substitute the limit.

\section{New methods for the early-time in-in integrals}
\label{othermethods}
\subsection{Different summation schemes}
\label{othersummation}
In this subsection, we provide different summation schemes, namely, Cesàro summation and Riesz summation,  which can also be applied to reorganize the target integral with immense convergence speed.
\subsubsection{Cesàro summation}
The Cesàro sum $ (C,\alpha) $ of an integral $ \int_{-\oo}^{\tau_{0}}f(\tau)d\tau $ is defined as \cite{hardydivseries},
\begin{align}
    \label{Cesarosum}
    \underset{(C,\alpha)}{\int_{-\oo}^{\tau_{0}}}f(\tau)d\tau  &\equiv\lim_{\tau\to-\infty}\dfrac{\int_{\tau_{0}}^{\tau}d\tau^{(\alpha)}\cdots\int_{\tau_{0}}^{\tau^{(2)}}d\tau^{(1)}\int_{\tau_{0}}^{\tau^{(1)}}d\tau^{(0)}f(\tau^{(0)})}{\int_{\tau_{0}}^{\tau}d\tau^{(\alpha)}\cdots\int_{\tau_{0}}^{\tau^{(2)}}d\tau^{(1)}\int_{\tau_{0}}^{\tau^{(1)}}d\tau^{(0)}\delta(\tau^{(0)}-\tau_{0})}\notag\\
	&=\lim_{\tau\to-\infty}-\dfrac{\alpha!}{(\tau-\tau_{0})^{\alpha}}\int_{\tau_{0}}^{\tau}d\tau^{(\alpha)}\cdots\int_{\tau_{0}}^{\tau^{(2)}}d\tau^{(1)}\int_{\tau_{0}}^{\tau^{(1)}}d\tau^{(0)}f(\tau^{(0)})~,
\end{align}
where $ \delta $ is the Dirac delta function. Since the Cesàro sum $ (C,\alpha) $ is compatible and equivalent to the Hölder sum $ (H,\alpha) $ \cite{hardydivseries}, then the value of the integral $ \int_{-\oo}^{\tau_{0}}f(\tau)d\tau $ with the the $ i\epsilon $ prescription is also equal to its $ (C,\alpha) $ sum for a large enough non-negative integer $ \alpha $. Like the case of the Hölder sum, if the non-oscillatory factor $ g(\tau) $ has the form $ g(\tau)\sim\tau^{n} $, then it  is summable by $ (C,\alpha) $ ($i.e.$ the limit in \eqref{Cesarosum} converges) if and only if $ \alpha>n $. More details of the Cesàro sum is in the Appendix \ref{H-C-Rconvergence}.

\subsubsection{Riesz summation}
The Riesz sum $ (R,\tau,\alpha) $ of an integral $ \int_{-\oo}^{\tau_{0}}f(\tau)d\tau $ is defined to be \cite{titchmarsh},
\begin{align}
    \label{Rieszsum}
    \underset{(R,\tau,\alpha)}{\int_{-\oo}^{\tau_{0}}}f(\tau)d\tau\equiv\lim_{\tau\to-\infty}\int_{\tau}^{\tau_{0}}\left(1-\dfrac{\tau'-\tau_{0}}{\tau-\tau_{0}}\right)^{\alpha}f(\tau')d\tau'~.
\end{align}
The Riesz sum $ (R,\tau,\alpha) $ is identical to the Cesàro sum $ (C,\alpha) $ because of the Fubini's Theorem,
\begin{align}
    &-\dfrac{\alpha!}{(\tau-\tau_{0})^{\alpha}}\int_{\tau_{0}}^{\tau}d\tau^{(\alpha)}\cdots\int_{\tau_{0}}^{\tau^{(2)}}d\tau^{(1)}\int_{\tau_{0}}^{\tau^{(1)}}d\tau^{(0)}f(\tau^{(0)})\notag\\
    =&-\dfrac{\alpha!}{(\tau-\tau_{0})^{\alpha}}\int_{\tau_{0}}^{\tau}d\tau^{(0)}f(\tau^{(0)})\int_{\tau^{
    (0)}}^{\tau}d\tau^{(\alpha)}\cdots\int_{\tau^{(0)}}^{\tau^{(2)}}d\tau^{(1)}\notag\\
    =&\int_{\tau}^{\tau_{0}}\left(1-\dfrac{\tau^{(0)}-\tau_{0}}{\tau-\tau_{0}}\right)^{\alpha}f(\tau^{(0)})d\tau^{(0)}~.
\end{align}
\noindent Therefore, the value of the integral $ \int_{-\oo}^{\tau_{0}}f(\tau)d\tau $ under the the $ i\epsilon $ prescription is also equal to its $ (R,\tau,\alpha) $ sum for a large enough non-negative integer $ \alpha $. However, the Riesz sum has only one layer of integral. Thus, we naturally expect that the computation time of the Riesz sum is shorter than that of the Cesàro sum. Similar to the Cesàro sum, if the non-oscillatory factor $ g(\tau) $ has the form $ g(\tau)\sim\tau^{n} $, then it is summable by  $ (R,\tau,\alpha) $ ($i.e.$ the limit in \eqref{Rieszsum} converges) if and only if $ \alpha>n $. The proof is given in Appendix \ref{H-C-Rconvergence}.

Although the Hölder sum and the Cesàro/Riesz sum are equivalent to each other and they all converge to the same value, their behavior at a finite early time cut-off are different, which is demonstrated in Appendix \ref{comparisonofsummationmethods}.

\subsubsection{Beta regulator method}

In Appendix \ref{H-C-Rconvergence}, we prove that the convergence speed of the Riesz sum is $ O(\tau^{-1}) $. Actually, the performance on the convergence speed can be further improved  by carefully choosing the weighted mean of Riesz sums. For example, we can slightly modify the integration as,
\begin{align}
    \label{modifiedRieszsum}
    \dfrac{1}{1-\frac{\alpha}{\alpha+1}}\lim_{\tau\to-\infty}\int_{\tau}^{\tau_{0}}\left[\left(1-\dfrac{\tau'-\tau_{0}}{\tau-\tau_{0}}\right)^{\alpha}-\frac{\alpha}{\alpha+1}\left(1-\dfrac{\tau'-\tau_{0}}{\tau-\tau_{0}}\right)^{\alpha+1}\right]f(\tau')d\tau'~~\text{ }(\alpha>n+1),
\end{align}
the above formula \eqref{modifiedRieszsum} converges to the desired value for sufficiently large $ \alpha $ with upgraded convergence speed $ O(\tau^{-2}) $. Follow the same strategy, a weighted mean formula with optimized convergence speed $ O(\tau^{-q}) $ reads as,
\begin{align}
    \label{generalmodifiedRieszsum}
    \lim_{\tau\to-\infty}\int_{\tau}^{\tau_{0}}\dfrac{\sum_{i=0}^{q-1}(-1)^{i}\binom{q-1}{i}\frac{\alpha}{\alpha+i}\left(1-\frac{\tau'-\tau_{0}}{\tau-\tau_{0}}\right)^{\alpha+i}}{\sum_{i=0}^{q-1}(-1)^{i}\binom{q-1}{i}\frac{\alpha}{\alpha+i}}f(\tau')d\tau'\qquad\text{ }(\alpha>n+q-1).
\end{align}
By putting the sum of the weights inside the integral with an appropriate index $p$, the numerical computation becomes more stable. Actually, the summation is nothing else but the regularized incomplete beta function $I_x(\alpha,q)$,
\begin{align}
    \label{rieszfactorandbeta}
    \dfrac{\sum_{i=0}^{q-1}(-1)^{i}\binom{q-1}{i}\frac{\alpha}{\alpha+i}\left(1-\frac{\tau'-\tau_{0}}{\tau-\tau_{0}}\right)^{\alpha+i}}{\sum_{i=0}^{q-1}(-1)^{i}\binom{q-1}{i}\frac{\alpha}{\alpha+i}}=I_{\frac{\tau-\tau'}{\tau-\tau_{0}}}(\alpha,q)~,
\end{align}
which can be easily evaluated with the built-in function in Mathematica.
In such sense, the integral with beta regulator can be defined as
\begin{align}
    \label{betaregulator}
    \underset{I_x(\alpha,q)}{\int_{-\oo}^{\tau_{0}}}f(\tau)d\tau\equiv\lim_{\tau\to-\infty}\int_{\tau}^{\tau_{0}}I_{\frac{\tau-\tau'}{\tau-\tau_{0}}}(\alpha,q)f(\tau')d\tau'~,
\end{align}
and it is one of the simplest methods to implement, since we only need to compute a 1D-integral of the original integrand with some extra regulator factor.

\subsection{Integral basis method} 
\label{integralbasis}
Besides the methods mentioned above, we have developed a new technique which separates the possible divergent part and treat different components independently. We will start from the simplest one-dimensional case to illustrate the basic idea, and then generalize it to more complicated cases.
\subsubsection{Single integral case}
\label{singleintegral}
As we have argued before, the behavior of the non-oscillatory factor can be well approximated by a polynomial $g(\tau)\sim\sum_{m=n_{1}}^{n_{2}}a_{m}\tau^{m} $ for some finite integers $n_{1} $, $ n_{2}$\footnote{Not necessarily an integer, but those non-integer power can also be fitted by one integer polynomial.}. Here we only focus on the early time divergence where the terms with negative power are suppressed. Hence, $ g(\tau) $ can be well approximated as $ \sum_{m=0}^{n}a_{m}\tau^{m} $ for some non-negative integer $ n $.
Let us denote the exact integrand with $ f(\tau) $ and its early time approximation as,
\begin{equation}{\label{earlytimeapp}
f^{\text{app}}(\tau)\equiv e^{\pm i\omega\tau}\sum_{m=0}^{n}c_{m}\tau^{m}}~.
\end{equation}
Now, we can decompose the integral into two terms using our integral basis,
\begin{align}
\label{decomposition}
\int_{-\oo^{\mp}}^{\tau_{\text{end}}}f(\tau)d\tau=\int_{-\oo^{\mp}}^{\tau_{\text{end}}}\left[f(\tau)-f^{\text{app}}(\tau)\right]d\tau+\int_{-\oo^{\mp}}^{\tau_{\text{end}}}f^{\text{app}}(\tau)d\tau~.
\end{align}
At the early stage, $ f(\tau)-f^{\text{app}}(\tau)\sim \mathcal{O}(\tau^{-1})\times e^{\pm i\omega\tau} $, which is self-convergent even without using $ i\e $-prescription, and can be easily evaluated numerically. 
For the other side, the remaining term can be integrated out analytically as the sum of the series of incomplete Gamma functions. More specifically,
\begin{align}\label{divergentterm}
 \int_{-\oo^{\mp}}^{\tau_{\text{end}}}f^{\text{app}}(\tau)d\tau=-\sum_{m=0}^{n}c_{m}(\mp i\omega)^{-m-1}\Gamma(m+1,\mp i\omega\tau_{\text{end}})~.
\end{align}
By extracting out the divergent part $f^{\text{app}} $, we reduce the computational errors to numerical ones that lie in the integration of $|f(\tau)-f^{\text{app}}(\tau)|$. 

\subsubsection{Generalization to multi-dimensional integrals}
The implementation for single integral can be extended to higher dimensional integral cases by performing the separation sequentially on each variable. To illustrate the idea, we consider a time-ordered two-dimensional integral,
\begin{align}
    \label{2-integral}
   \int_{-\infty}^{\tau_{0}}\int_{-\infty}^{\tau_{1}}f_{1}(\tau_{1})f_{2}(\tau_{2})d\tau_{2}d\tau_{1}~,
\end{align}
where the functions $ f_{1} $ and $ f_{2} $ both have the early time approximation in the form of \eqref{earlytimeapp}. By adopting the integral basis for single layer integration in Section~\ref{singleintegral}, we can integrate over $ \tau_{2} $ to obtain a function of $ \tau_{1} $ which reads as,
\begin{align}
    \label{upperlimitfunction}
    F_{2}(\tau_{1})\equiv\int_{-\infty}^{\tau_{1}}f_{2}(\tau_{2})d\tau_{2}~.
\end{align}
The early time behaviour of $ F_{2}(\tau_{1}) $ follows the analytical integration in \eqref{divergentterm}, 
\begin{align}
    F_{2}(\tau_{1})\approx\int_{-\oo^{\mp}}^{\tau_{1}}f_{1}^{\text{app}}(\tau)d\tau=-\sum_{m=0}^{n}c_{m}(\mp i\omega)^{-m-1}\Gamma(m+1,\mp i\omega\tau_{1})~.
\end{align}
Since the incomplete Gamma function with the first argument being a positive integer has the form of the RHS of \eqref{earlytimeapp}, $ F_{2}(\tau_{1}) $ also takes the early time approximation of this form, so does $ f_{1}(\tau_{1})F_{2}(\tau_{1}) $. Furthermore, we can use the one-dimension integral procedure one more time on
\begin{align}
    \int_{-\infty}^{\tau_{0}}f_{1}(\tau_{1})F_{2}(\tau_{1})d\tau_{1}~,
\end{align}
to obtain the final result of \eqref{2-integral}. More technical details of the implementation of the procedure presented in this section can be found in the Appendix \ref{technicaldetails}.

\subsection{Partition-Extrapolation methods}
\label{partitionmethods}
The Partition-Extrapolation methods are highly efficient in calculating the oscillatory infinite-range integrals that we usually encounter. In the electromagnetic engineering area, these methods have already been applied to calculate the tail of the Sommerfeld integrals with great success \cite{pemethodsreview,pemethodsreview2}. We state such a powerful method in this section instead of under the discussion of Section~\ref{knownmethods} for known methods, because it is rarely used for calculating the cosmological correlators. Considering their outstanding performance, these methods should deserve the most attention. 

A pedagogical review including efficiency comparison of PE methods together with pseudocodes have been already provided in \cite{pemethodsreview,pemethodsreview2}. In this subsection, we summarize the most important points from these papers and focus on two best PE methods: the Levin-Sidi and Mosig-Michalski algorithms. Another algorithm that also deserves attention is the Shanks-Wynn algorithm \cite{pemethodsreview2}, which will not be included here. 

The principal idea of partition-extrapolation methods is to turn the problem of calculating the integral
\begin{align}
    \int_{-\oo}^{\tau_{0}}f(\tau)d\tau~,
\end{align}
into the computation of the series
\begin{align}
    \label{integralseries}
    S=\sum_{i=0}^{\infty}u_{i},\text{ }u_{i}= \int_{\xi_{i+1}}^{\xi_{i}}f(\tau)d\tau~,
\end{align}
where $ \{\xi_{i}\}_{i=0}^{\infty} $ is a monotonically decreasing sequence diverging to $ -\infty $ with $ \xi_{0}=\tau_{0} $. This sequence is usually called "break points", which explains the word "partition" in the name of PE methods. We also note that the value of $ S $ can also be understood as the limit of the sequence of partial sums,
\begin{align}
    S_{n}=\sum_{i=0}^{n}u_{i}~.
\end{align}
Then the result of integration can be obtained by applying sequence convergence acceleration methods to the sequence $ \{S_{n}\} $. 

Levin-Sidi and Mosig-Michalski algorithms are mainly used to deal with the integrals with integrand of the type
\begin{align}
    \label{integrandform}
    f(\tau)=g(\tau)p(\tau)~,
\end{align}
that $ g(\tau) $ has the early-time approximation like,
\begin{align}
    \label{nonosform}
    g(\tau)\sim\dfrac{e^{\zeta\tau}}{(-\tau)^{\alpha}}\sum_{j=0}^{\infty}a_{j}\tau^{-j}\qquad\text{ }\tau\to-\infty~,
\end{align}
where $ \alpha $ is real, $ a_{0}\neq0 $ and $ \zeta\geq0 $; while $ p(\tau) $ is a periodic function with half-period $ q $ satisfying
\begin{align}
    \label{osform}
    p(\tau+q)=-p(\tau)~.
\end{align}
We have already argued that  our target integrands can be classified into this type satisfying conditions with $ \zeta=0 $ and $ p(\tau)=e^{i\omega\tau} $. Therefore, it is possible to apply the Levin-Sidi and Mosig-Michalski algorithms to the in-in integrals. 

In general, the convergence of a sequence could be classified based on
\begin{align}
    \lambda=\lim_{n\to\infty}\dfrac{r_{n+1}}{r_{n}}~,
\end{align}
where $ r_{n} $ is the remainder which is defined as $ r_{n}\equiv S-S_{n} $. The convergence is called ``linear'' if $ |\lambda|<1 $, ``logarithmic'' if $ \lambda=1 $, and ``hyperlinear'' if $ \lambda=0 $.\cite{Weniger:1989rea} If $ \lambda>0 $, the sequence is ``asymptotically monotone'' and if $ \lambda<0 $, the sequence is ``alternating''.\cite{pemethodsreview2}

It turns out that the sequence convergence acceleration methods work most efficiently with alternating sequences. With the integrands satisfying \eqref{integrandform}, we can turn $ S_{n} $ into alternating sequence by choosing
\begin{align}
    \label{breakpoints}
    \xi_{n}=\tau_{0}-nq,\qquad(n=0,1,2,\cdots)~.
\end{align}
Indeed, it is readily proved that \cite{pemethodsreview,pemethodsreview2} in this case, the remainder will take the form
\begin{align}
    \label{remainderform}
    r_{n}=\omega_{n}\sum_{j=0}^{\infty}c_{j}\xi_{n}^{-j}\qquad n\to\infty,
\end{align}
where $ c_{0}\neq0 $ and $ \omega_{n} $ are the "remainder estimates", which in this case have the expression
\begin{align}
    \label{exactremainderestimate}
    \omega_{n}=(-1)^{n+1}\dfrac{e^{-n\zeta q}}{(-\xi_{n})^{\alpha}}~.
\end{align}

\subsubsection{The Levin-Sidi method}

The idea of Levin-Sidi algorithm is based on a simple mechanism. Firstly, we truncate the sum in \eqref{remainderform} to $ j=k-1 $. Then, by knowing the value of $ S_{n} $ and $ \omega_{n} $ at $ k+1 $ different values of $ n $, we can set up a linear system of $ k+1 $ equations to solve for the unknowns $ \{S,c_{0},\cdots,c_{k-1}\} $ by the truncated \eqref{remainderform}. The value of $ S $ obtained from solving this system is our estimate for the integral value. 

The value of $ S_{n} $ could be obtained by direct integration. The way we determine the value of $ \omega_{n} $ gives rise to different variants of the Levin-Sidi method. If the exact expression $ \eqref{exactremainderestimate} $ is used, we have the $ a $-variant. However, this requires prior knowledge of the degree $ \alpha $, or at least, $ \alpha $ have to be determined through regression. 

An alternative way is to estimate the value of $ \omega_{n} $ and Levin has derived some estimations which produce satisfactory result \cite{levintranspaper,smithfordd-trans}
\begin{align}
    \label{levintransformation}
    \omega_{n}=\begin{cases}
    u_{n}\text{ or }u_{n+1} &(\text{$t-$ transformation or $d-$ transformation})\\
    \xi_{n}u_{n} &(\text{$u-$ transformation})\\
    \dfrac{u_{n}u_{n+1}}{u_{n}-u_{n+1}} &(\text{$v-$ transformation})~.
\end{cases}
\end{align}
These choices of remainder estimate can accelerate a broad class of series. The $ t- $ transformation can accelerate linear and alternating series, but not logarithmic series. Meanwhile, the $ u- $ and $ v- $ transformations can accelerate both linear and logarithmic series. The series \eqref{integralseries} ($ \zeta=0 $) corresponding to the integral we cared is alternating. Thus, all four transformations can be applied to compute the result in this case.

We can understand qualitatively why these estimations work in the case we are interested. Indeed, from \eqref{remainderform} and \eqref{exactremainderestimate}, we have
\begin{align}
    u_{n}&=r_{n-1}-r_{n}\nonumber\\
    &=(-1)^{n}\sum_{j=0}^{\infty}(-1)^{j}c_{j}\left[(-\xi_{n})^{-(\alpha+j)}+(-\xi_{n}+q)^{-(\alpha+j)}\right]\notag\\
    &=-\dfrac{2c_{0}(-1)^{n+1}}{(-\xi_{n})^{\alpha}}+\mathcal{O}\left((-\xi_{n})^{-(\alpha+1)}\right) \qquad n\to\infty~.
\end{align}
The second line is obtained by Taylor expansion of the second term. Consequently, the leading dependence of $ u_{n} $ on $ \xi_{n} $ only differs from exact $ \omega_{n} $ (Eq.\eqref{exactremainderestimate}) by a constant factor independent of $ n $, which can be absorbed into the unknowns $ c_{j} $ of the linear system. Thus, we can also use $ u_{n} $ as an estimate for $ \omega_{n} $. The same argument also holds for the $ d $- and $v$- transformations. Following this spirit, it is also possible to use $ f(\xi_{n}) $ as the estimation for $ \omega_{n} $ since it also has the asymptotic leading dependence $ f(\xi_{n})\sim(-1)^{n}(-\xi_{n})^{-\alpha} $. For the $  u $-transformation, the corresponding leading dependence is $ \xi_{n}u_{n}\sim(-1)^{n}(-\xi_n)^{-(\a-1)}$ and the expression of the exact remainder \eqref{remainderform} under the transformation is simply the special case when $ c_{0}=0 $. Thus, the $ u $-transformation does not remove any meaningful unknown from the linear system to be solved for obtaining an estimation of $ S $.

The linear system that we need to solve is very similar to the linear system in the polynomial interpolation problem, where the solution for each unknown is obtained from the Newton divided difference formula. We also have a similar recursive algorithm to obtain the result in the problem that we are concerning, which is invented by Sidi \cite{Sidioriginalpaper}, dubbed W-algorithm. The principal formula of the W-algorithm for the estimate of $ S $ is
\begin{align}
    S^{(k)}=\dfrac{A_{0}^{(k)}}{B_{0}^{(k)}},\text{ }A_{n}^{(k)}=\delta^{k}(S_{n}/\omega_{n}),\text{ }B_{n}^{(k)}=\delta^{k}(1/\omega_{n})~,
\end{align}
where $ \delta $ is the Newton divided difference of variable $ \xi^{-1} $, defined recursively for a sequence $ \{R_{n}\} $ as
\begin{align}
    \delta^{0}(R_{n})&=R_{n}\\
    \delta^{k+1}(R_{n})&=\dfrac{\delta^{k}(R_{n+1})-\delta^{k}(R_{n})}{\xi_{n+k+1}^{-1}-\xi_{n}^{-1}}.
\end{align}
We refer the readers to \cite{pemethodsreview2} for a pseudocode implementing this algorithm and to \cite{sidiintegral1,sidiintegral2} for a rigorous analysis.

\subsubsection{The Mosig-Michalski method}

The Mosig-Michalski method transforms the sequence $ \{S_{n}\} $ into a new sequence $ \{S'_{n}\} $ which converges to the same limit at faster rate than the original one. Denote the remainder of the new sequence to be $ r'_{n}\equiv S-S'_{n} $. By "converging faster", we mean that the following condition is satisfied
\begin{align}
    \label{acceleratingcondition}
    \dfrac{|r'_{n}|}{|r_{n}|}=O(\xi_{n}^{-\mu}),\qquad\mu>0.
\end{align}
We consider the sequence transformation which takes the weighted average of two consecutive elements
\begin{align}
    S'_{n}=\dfrac{S_{n+1}-\eta_{n}S_{n}}{1-\eta_{n}},\qquad(\eta_{n}\neq1).
\end{align}
The Mosig-Michalski method uses a particular choice of $ \eta_{n} $ to achieve condition~\eqref{acceleratingcondition}. More specifically, the ratio of two reminders can be expressed as
\begin{align}
    \label{newremainder}
    \dfrac{r'_{n}}{r_{n}}\equiv \dfrac{S-S'_{n}}{S-S_{n}}=\dfrac{r_{n+1}/r_{n}-\eta_{n}}{1-\eta_{n}},
\end{align}
the remainder can be cancelled completely with the choice $ \eta_n=r_{n+1}/r_{n} $. With lacking of precise information about $ r_n$, we are only able to estimate its value. Nevertheless, we can still pick particular $ \eta_{n} $ which is extremely close to $ r_{n+1}/r_{n} $. By substituting $ r_{n} $ from \eqref{remainderform} and performing Taylor expansion, we find
\begin{align}
    \label{remainderratio}
    \dfrac{r_{n+1}}{r_{n}}&=\dfrac{\omega_{n+1}}{\omega_{n}}\dfrac{\sum_{j=0}^{\infty}c_{j}(\xi_{n}-q)^{-j}}{\sum_{j=0}^{\infty}c_{j}\xi_{n}^{-j}}\nonumber\\&=\dfrac{\omega_{n+1}}{\omega_{n}}\left(1+\dfrac{c_{1}}{c_{0}}\xi_{n}^{-1}+O(\xi_{n}^{-2})\right)\left(1-\dfrac{c_{1}}{c_{0}}\xi_{n}^{-1}+O(\xi_{n}^{-2})\right)\notag\\
    &=\dfrac{\omega_{n+1}}{\omega_{n}}+O(\xi_{n}^{-2})~,
\end{align}
here we used $ \omega_{n+1}/\omega_{n}=O(\xi_{n}^{0}) $ to bring $ O(\xi_{n}^{-2}) $ outside of the bracket. Therefore, a reasonable choice is $ \eta_{n}=\omega_{n+1}/\omega_{n} $ which can achieve the numerator in \eqref{newremainder} is $ O(\xi_{n}^{-2}) $. In this case, by substituting $ \omega_{n} $ from \eqref{exactremainderestimate} with $ \zeta=0 $, we can get
\begin{align}
    \label{1-eta}
    1-\eta_{n}=(-1)^{n+1}\dfrac{(-\xi_{n}+q)^{-\alpha}+(-\xi_{n})^{-\alpha}}{(-\xi_{n})^{-\alpha}}=O(\xi_{n}^0).
\end{align}
As a result, the condition \eqref{acceleratingcondition} is satisfied with $ \mu=2 $. With the exact expression of $ \omega_{n} $ Eq.\eqref{exactremainderestimate}, the desired weights $ \eta_{n} $ can be computed analytically.

Alternatively, we may also estimate the value of $ \omega_{n} $ up to a constant factor (since we are calculating ratio of $ \omega_{n} $-s) with the $ t $-, $ d $-, $ v $- transformations of Levin \eqref{levintransformation}.\cite{pemethodsreview} For example, consider the case of the $ t $-transformation
\begin{align}
    \dfrac{u_{n+1}}{u_{n}}&=\dfrac{r_{n}-r_{n+1}}{r_{n-1}-r_{n}}=\dfrac{r_{n+1}}{r_{n}}\dfrac{r_{n}/r_{n+1}-1}{r_{n-1}/r_{n}-1}=\dfrac{r_{n+1}}{r_{n}}\dfrac{\omega_{n}/\omega_{n+1}-1+O(\xi_{n}^{-2})}{\omega_{n-1}/\omega_{n}-1+O(\xi_{n}^{-2})}~,
\end{align}
 note that
\begin{align}
    \omega_{n}/\omega_{n+1}&=-\dfrac{(-\xi_{n})^{-\alpha}}{(-\xi_{n}+q)^{-\alpha}}=-1-\alpha\dfrac{q}{-\xi_{n}}+O(\xi_{n}^{-2}),\notag\\
    \omega_{n-1}/\omega_{n}&=-\dfrac{(-\xi_{n}-q)^{-\alpha}}{(-\xi_{n})^{-\alpha}}=-1-\alpha\dfrac{q}{-\xi_{n}}+O(\xi_{n}^{-2})~,
\end{align}
then
\begin{align}
    \dfrac{u_{n+1}}{u_{n}}&=\dfrac{r_{n+1}}{r_{n}}\left[1+O(\xi_{n}^{-2})\right]=\dfrac{r_{n+1}}{r_{n}}+O(\xi_{n}^{-2}).
\end{align}
This implies $ r_{n+1}/r_{n}=u_{n+1}/u_{n}+O(\xi_{n}^{-2}) $, which is analogous to \eqref{remainderratio}. By choosing $ \eta_{n}=u_{n+1}/u_{n} $, we can prove in a similar way to \eqref{1-eta} that $ 1-\eta_{n}=O(\xi_{n}^{0}) $. Thus, with the $ t $-transformation, we still achieve \eqref{acceleratingcondition} with $ \mu=2 $.

The $ u $-transformation, though still works, is not preferred since it scales with $ (-\xi_{n})^{-\alpha+1} $ instead of $ (-\xi_{n})^{-\alpha} $ like the exact $ \omega_{n} $. Because of this, it only achieves $ \mu=1 $. This can be seen from
\begin{align}
    \dfrac{\xi_{n+1}}{\xi_{n}}\dfrac{u_{n+1}}{u_{n}}&=\left[1+\dfrac{q}{-\xi_{n}}\right]\dfrac{r_{n+1}}{r_{n}}\left[1+O(\xi_{n}^{-2})\right]=\dfrac{r_{n+1}}{r_{n}}\left[1+O(\xi_{n}^{-1})\right].
\end{align}
The process of taking weighted average can be performed iteratively to generate a list of sequences $ \{\{S_{n}^{(0)}\},\{S_{n}^{(1)}\},\cdots,\{S_{n}^{(k)}\}\} $ where $ \{S_{n}^{(0)}\}\equiv\{S_{n}\} $ and $ \{S_{n}^{(1)}\} $ is obtained from the procedure described above. For further iterations with $ m\geq1 $
\begin{align}
    S^{(m+1)}_{n}=\dfrac{S^{(m)}_{n+1}-\eta_{n}^{(m)}S^{(m)}_{n}}{1-\eta_{n}^{(m)}} \qquad(\eta_{n}^{(m)}\neq1)~,
\end{align}
we need to choose the weights $ \eta_{n}^{(m)} $ such that it approximates $ r_{n+1}^{(m)}/r_{n}^{(m)} $. From \eqref{acceleratingcondition}, the remainder $ r'_{n} $ also has the form of \eqref{remainderform} with $ \omega_{n}'=\omega_{n}\xi_{n}^{-\mu} $. This leads to the choice of weights
\begin{align}
    \eta_{n}^{(m)}=\eta_{n}^{(0)}\left(\dfrac{\xi_{n}}{\xi_{n+1}}\right)^{\mu}~.
\end{align}
We refer the readers to \cite{pemethodsreview2} for a pseudocode implementing this algorithm.

\section{Application to typical integrals}
\label{comparison} 
\subsection{Example 1: the integral with analytical expression }
\begin{figure}[htp]
    \centering
   \includegraphics[width=0.75\textwidth]{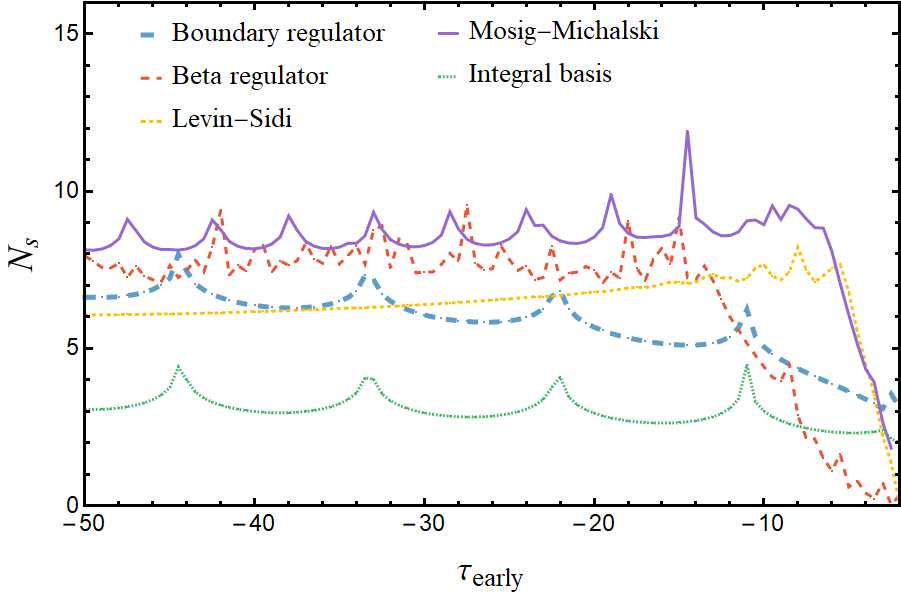}
    \caption{Comparison of the convergence ability of different methods\protect\footnotemark.}
    \label{fig:analytical_significantdigit}
\end{figure}

\begin{figure}[H]
    \centering
   \includegraphics[width=0.75\textwidth]{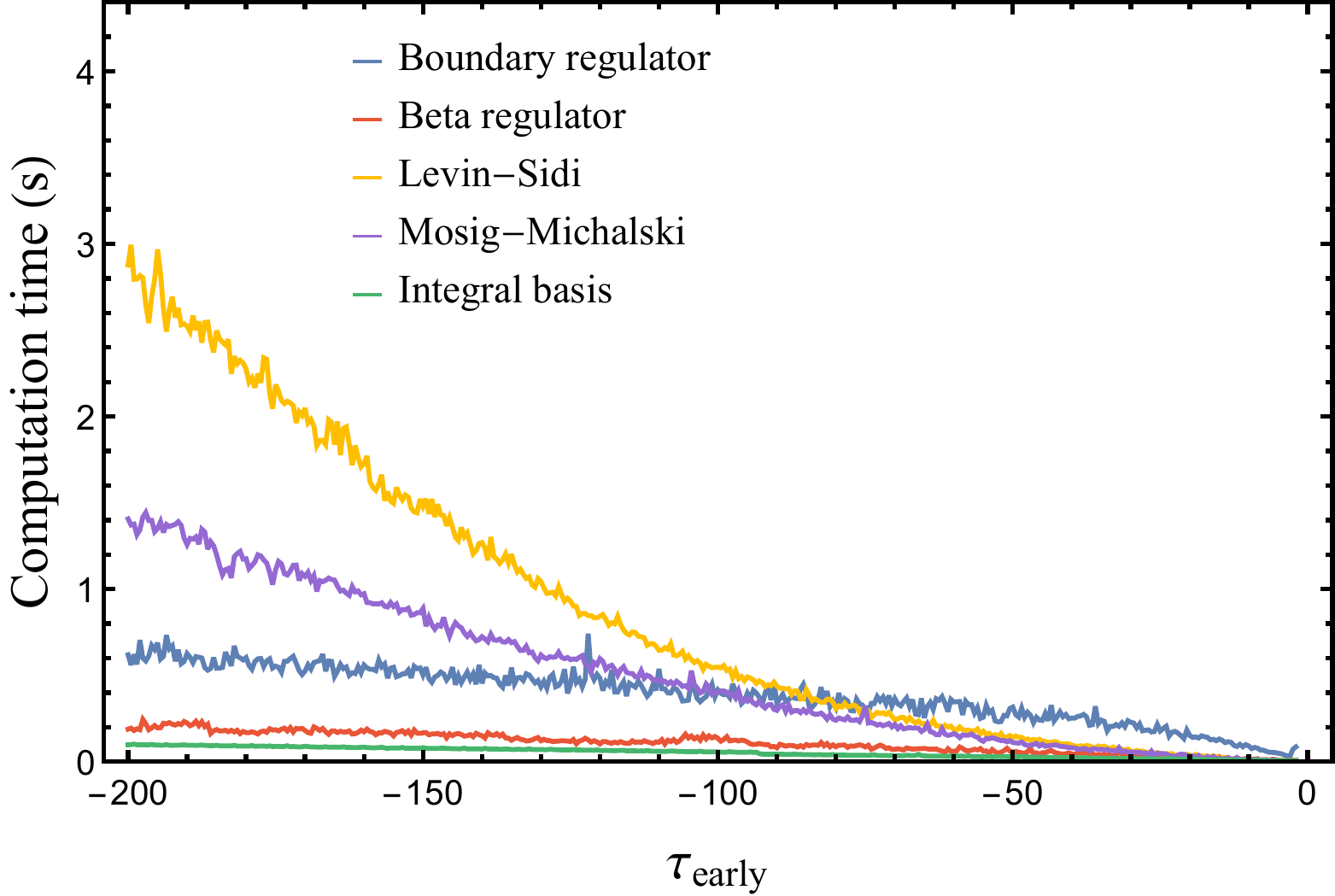}
  
    \caption{The dependence of the computation time via different methods on the early time cut-off $\tau_{\text{early}}$~.}
    \label{fig:analyticalcomputationtimes}
\end{figure}

As a warm-up exercise, and for comparing the performance of different methods including their accuracy, time consumption as well as the convergence speed, we first implement those typical methods to evaluate a simply integral which can be easily solved analytically. To be more specific, let us consider the  below integral which may appear in the bispectrum with one graviton external leg
\begin{equation}
\label{integrandI}  
I =\mathfrak{R}\left[ - \int^0_{-\oo} i\frac{d\eta}{\eta^2} (1-ik_1 \eta)(1-ik_2 \eta)(1-ik_3 \eta) e^{ik_t \eta} \right]~,
\end{equation}
where $ k_{t}\equiv k_{1}+k_{2}+k_{3} $,  the non-oscillatory prefactor is power law divergent at the far past infinity. Besides, the integrand also face the IR divergence at $ \eta=0 $. This does not pose any problem to the calculations, 
because the divergence is purely imaginary with a suitable choice of contour, while the final result will just take  the real part \cite{maldacena2003non}. In this paper, since the main purpose we are concerning about is to deal with the nonphysical early time divergence, then we will only  calculate the below toy integral
\begin{align}
\label{analyticaltoyI}
    I_{1}=\int^{-1}_{-\infty}-i\frac{d\eta}{\eta^2} (1-ik_1 \eta)(1-ik_2 \eta)(1-ik_3 \eta) e^{ik_t \eta}~.
\end{align}
\begin{figure}[htp]
\centering
\includegraphics[width=15.5cm]{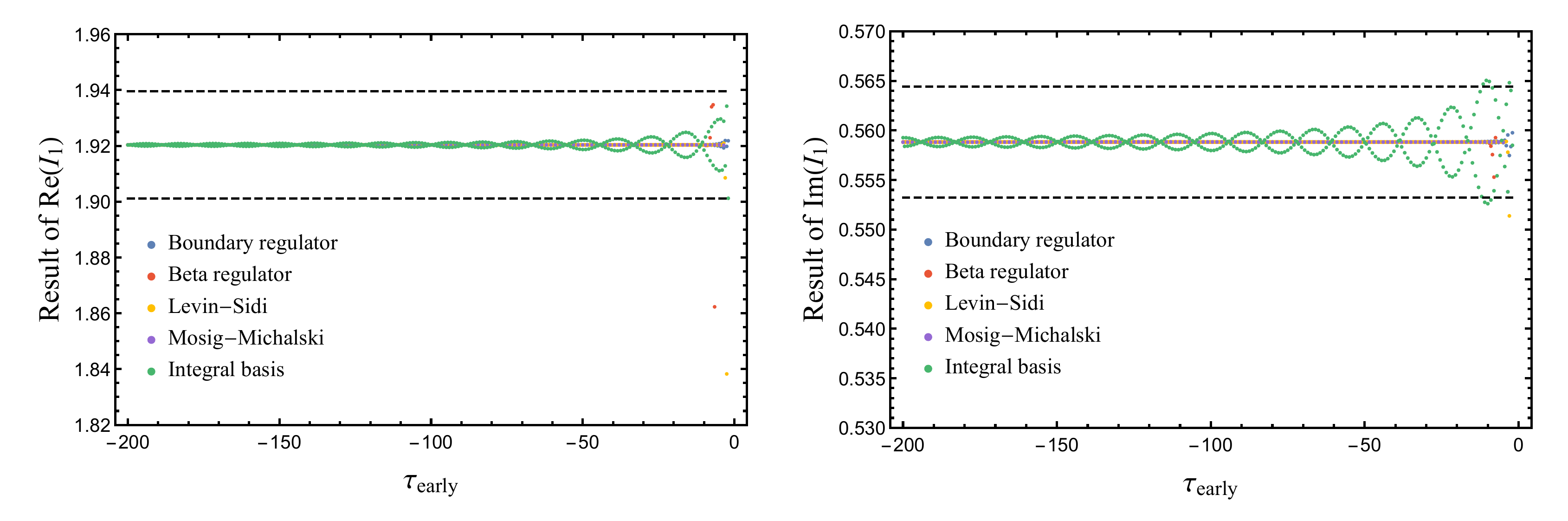}
\caption{The dependence of integration value via different methods on the early time cut-off $\tau_{\text{early}}$. The dashed line represents the 1\% error boundary.}
\label{analyticalconvergence}
\end{figure}
In our discussion, five typical methods with satisfactory performance are included. The Boundary regulator method is implemented with the minimal number of integrations by part (in this case $ p=1 $, so $ n=2 $). We take the Boundary Regulator with $ (\alpha,q)=(20,16) $ as the representative of this kind of summation methods, by considering its best performance and readers can find more details in Appendix \ref{comparisonofsummationmethods} about the comparison of these different summation methods. The Levin-Sidi and Mosig-Michalski method are implemented with the $ d- $ variant. Damping factor method is excluded at this moment, because its intrinsic error and damping factor $\beta$ should be chosen carefully to balance the efficiency and the accuracy. Readers can find more details provided in Appendix \ref{DFconv}. The numerical results of different typical methods as a function of early time cut-off $\tau_{\text{early}}$ are summarized in the Fig.~\ref{analyticalconvergence}, where black dashed line represents the $1\%$ error boundary. All of our calculations are conducted by Mathematica, in which the precision goals and the working precision of internal calculations are set to default values (see \cite{mathematicaprecision}).
As indicated in the figure, all methods converge to the same exact result with very high convergence ability. To better illustrate their accuracy together with convergence speed, we define the number of significant digits $N_s$, that can evaluate how close the numerical results are to the standard value
\begin{align}
    N_s\equiv-\log_{10}\left(\middle|\frac{I_N-I_s}{I_s}\middle|\right)~,
\end{align}
where $ I_{N} $, $ I_{s} $ are the numerical result obtained from different methods and the standard value of the integral that we refer to, respectively.
We compute the real part of $I_1$ with the result obtained by the Wick-rotation method being the standard value, shown in Fig.~\ref{fig:analytical_significantdigit}.
\footnotetext{Note that the number of significant digits of every method is bounded by the default precision setting of Mathematica for NDSolve and NIntegrate (roughly 8 digits).}
As the Figure shows, the PE methods including both the Mosig-Michalski and the Levin-Sidi have impressive performance in the convergence speed. Only after a small integration region $\Delta\tau\sim\mathcal{O}(1)$, the numerical result quickly converges to the standard value with extremely high accuracy (can be up to 8 significant digits). In contrast, other methods exhibit slightly lower but also acceptable convergence speed. Besides, another important thing in practical implementation is the computation time consumption. We compare the computation time of different methods given a certain early time cut-off $\tau_{\text{early}}$, and the result is summarized in the Fig.~\ref{fig:analyticalcomputationtimes}. Our technique, the Integral basis method manifests its great advantage in computation time consumption due to the fact that the hardest divergent part has already been separated out and mimicked by some incomplete Gamma functions which are easy to evaluate.
 For the convenience of readers, we summarize the performance of different methods in the Table.\ref{ranking1}~.
\begin{table}[htp]
\centering
\begin{tabular}{ |p{1.5cm}||p{4.5cm}|p{4.5cm}|  }
 \hline
 \multicolumn{3}{|c|}{Performance of different typical methods} \\
 \hline
Ranking&~~~~~Convergence speed &~~~~~Computation time\\
 \hline
 ~~~~~1&Mosig-Michalski&Integral basis\\
 ~~~~~2&Levin-Sidi~~($\approx1$)&Beta regulator\\
 ~~~~~3&Beta regulator&Mosig-Michalski\\
 ~~~~~4& Boundary regulator&Levin-Sidi\\
 ~~~~~5&Integral basis&Boundary regulator\\
 \hline
\end{tabular}
\caption{The performance ranking of different methods.}
\label{ranking1}
\end{table}
\subsection{Example 2: the integral with numerical mode function}
To further assess their application and show their advantages, we choose another integral in numerical form where analytical Wick rotation method is not possible. For example, the one appeared in the evaluation of 3-point functions of a featured-potential inflationary model which introduces a step into the slow-roll potential like \cite{chen2007large}
\begin{equation}
    V(\phi)=\frac{1}{2}m^2\phi^2\left[1+c\tanh\left(\frac{\phi-\phi_s}{d}\right)\right]~,
\end{equation}
where the step locates at $\phi_s$. By solving the equation of motion for scalar perturbations and the Mukhanov equation in conformal time \cite{Mukhanov:1985rz}, one can obtain the numerical-form integrands of the three-point correlation function. The parameters of the model are chosen as $(c,d,\phi_{s})=(0.002,0.02,15.86M_{p}) $. The initial conditions and the unit of conformal time are chosen such that the step and the horizon crossing of the mode $ k=1 $ occur around $ \tau=-1 $ and the inflation ends around $ \tau=0 $. 

Without loss of generality, we pick up one  integral $I_{2}$ from the integrals consist of the three-point correlation function,
\begin{align}
\label{FMtoyintegral}
I_{2} = -2i\int_{\tau_0}^{\tau_{end}}&d\tau\epsilon^2 a^2 u_{k_1}^{*}(\tau)u_{k_2}^{*}(\tau)u_{k_3}^{*}(\tau),
\end{align}
\begin{figure}[htp]
\centering
\includegraphics[width=15.5cm]{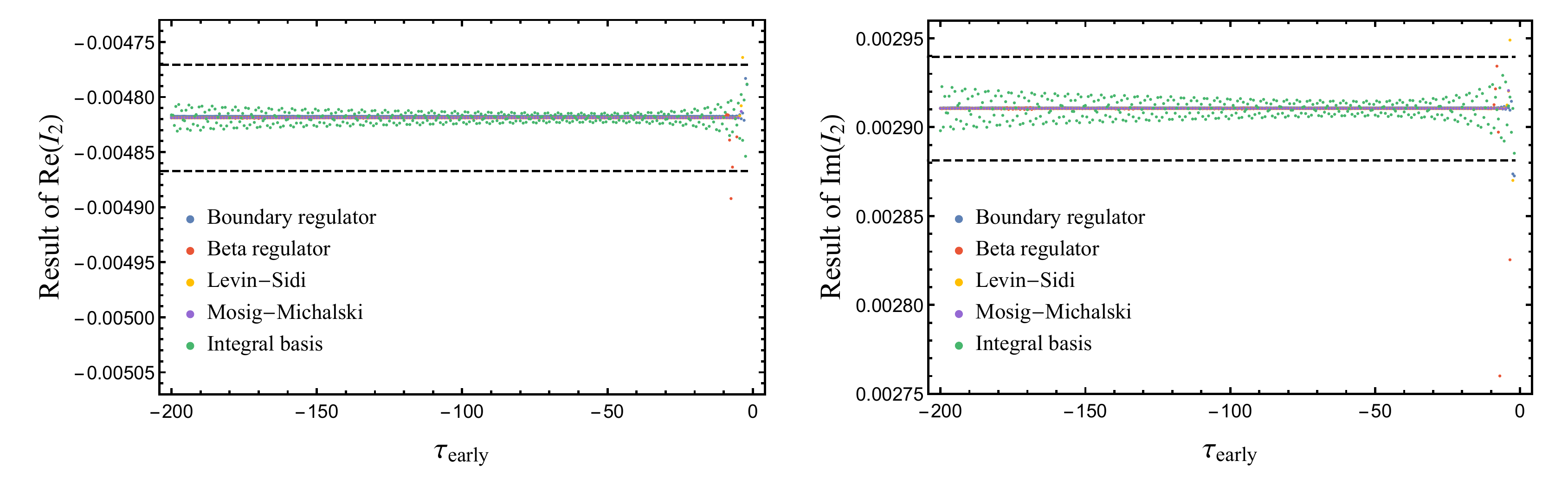}
\caption{The dependence of the integration value via different methods on the early time cut-off $\tau_{\text{early}}$ . The dashed line represents the 1\% error boundary.}
\label{fig:numerical_convergence}
\end{figure}
where the integrand is in the numerical form. The numerical integration is evaluated to the early time cut-off $ \tau_{\text{early}} $ and the dependence of the result on the cutoff value by different methods are summarized in Fig.~\ref{fig:numerical_convergence}, where we compute the real part of $I_2$. Shown clearly by the figure, all methods quickly converge to the same value.
\begin{figure}[htp]
    \centering
   \includegraphics[width=0.75\textwidth]{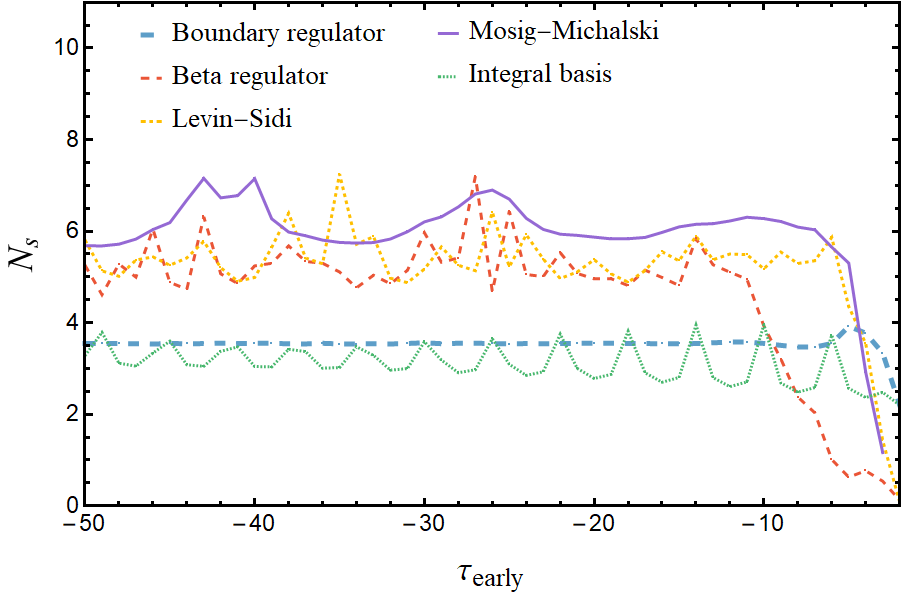}
  
    \caption{Comparison of the convergence ability of different methods.}
    \label{fig:numerical_digits}
\end{figure}

Unlike the Example 1 where we are able to find an analytical standard value from the wick rotation as the reference, the standard is absent here. Based on the experience of previous application, we choose the value obtained from the Mosig-Michalski method at large $\tau_\text{early}$ as the standard value $I_s$ to plot Fig.~\ref{fig:numerical_digits}, which shows the convergence ability of different typical approaches. All methods converge to their own desired value at very high speed, and the relative difference between the convergence values obtained by different methods is extremely small. Nevertheless, any prior assumptions about standard value inevitably introduce the bias, so we do not rank the accuracy or convergence speed of different methods here, which may be inappropriate and unfair.


\begin{figure}[H]
    \centering
   \includegraphics[width=0.75\textwidth]{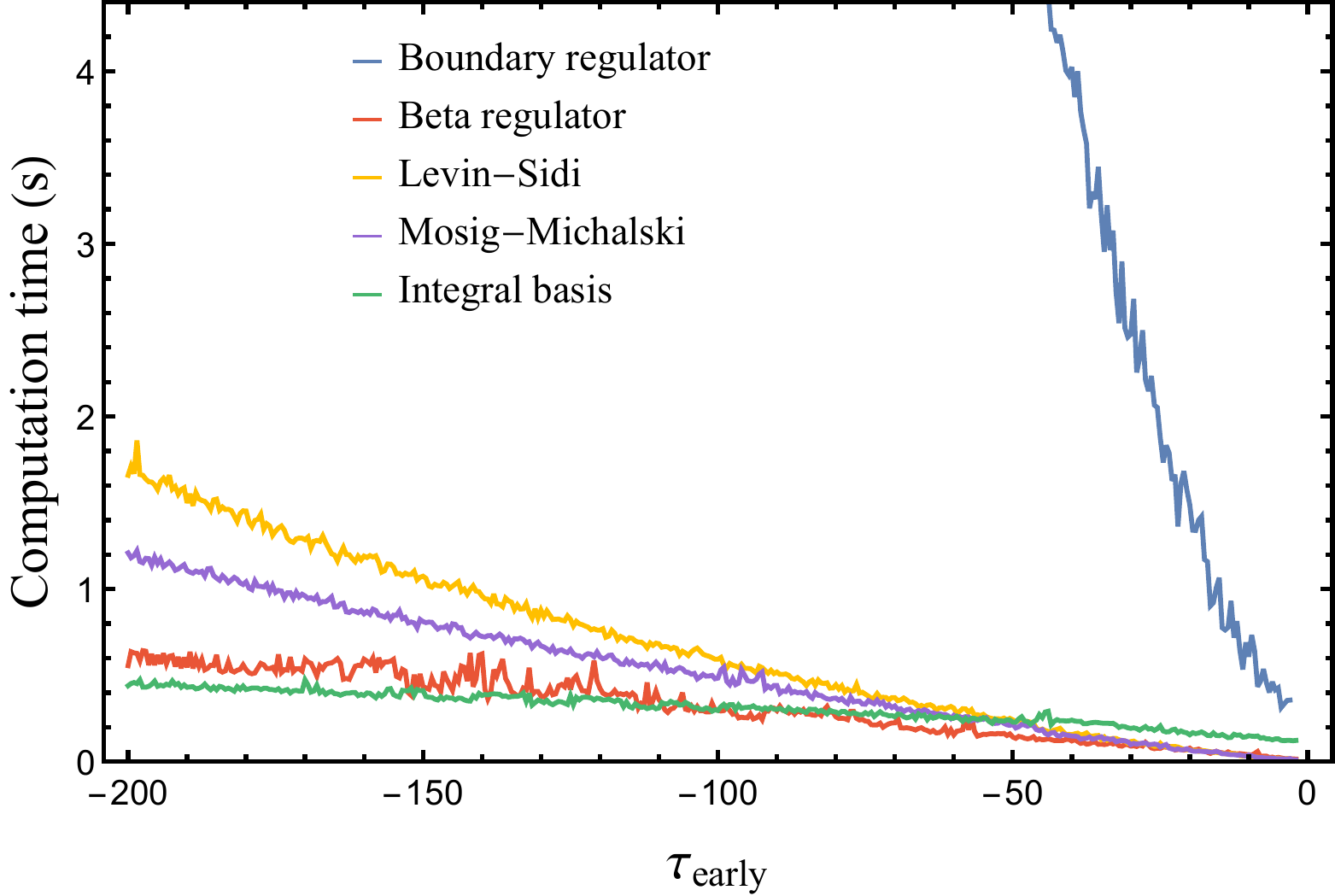}
  
    \caption{The dependence of the computation time via different methods on the early time cut-off $\tau_{\text{early}}$.}
    \label{fig:numericalcomputationtimes}
\end{figure}
In Fig.~\ref{fig:numericalcomputationtimes}, we summarized computation time consumption of different methods as a function of cut-off $\tau_\text{early}$. As indicated in the figure, the integral basis method again shows its superiority in this aspect, due to the powerful ability dealing with the divergent part of the integral. On the other hand, the boundary regulator method turns out to be much more time-consuming as it takes time to compute and call the numerical derivatives of the integrand during integrating by parts. For the convenience of readers, we also summarize the performance of different methods in the Table. \ref{ranking2}.
\begin{table}[htp]
\centering
\begin{tabular}{ |p{1.5cm}||p{4.5cm}|  }
 \hline
 \multicolumn{2}{|c|}{Performance of different typical methods} \\
 \hline
Ranking&~~~~Computation time\\
 \hline
 ~~~~~1&Integral basis\\
 ~~~~~2&Beta regulator\\
 ~~~~~3&Mosig-Michalski\\
 ~~~~~4& Levin-Sidi\\
 ~~~~~5& Boundary regulator\\
 \hline
\end{tabular}
\caption{The performance ranking of different methods.}
\label{ranking2}
\end{table}
\section{Conclusions} 
\label{conclusions}
Considering the situation that most of integrands in the correlation functions do not possess an analytical expression and they are highly oscillatory and divergent at the early time, we see the necessity of developing numerical techniques that can effectively suppress such divergence in a short time period to give the integration output.
In our work, we develop some new numerical techniques and compare the performance of different methods that can be applied to the computation of correlation functions in the in-in formalism. 

We started by a brief review of different techniques previously used to compute cosmological correlator (including Damping Factor, Boundary Regulator and Hölder Summation), and then introduced other summation-based methods (including Cesàro Summation, Riesz Summation and Beta Regulator), also the new numerical technique Integral Basis, and reviewed the Partition-Extrapolation methods (including Levin-Sidi and Mosig-Michalski). We have proved the convergence condition of Hölder, Cesàro/Riesz summations in the case where the early time integrand is in the form of $\tau^n e^{ik\tau}$ which is aligned with the early time mode functions solved in Bunch-Davies vacuum condition. Additionally, we have found and proved that the asymptotic convergence speed could be increased to arbitrary power by taking the weighted mean of Riesz sums and derived an explicit formula of regularized incomplete beta function which works.

We have developed and presented detailed description of Integral Basis to employ the numerical integration and reduce the computation errors to only numerical errors. By decomposing the integrand into early time analytical approximation and numerical components, we obtained an effective convergence in both numerical and analytical evaluation. We have also generalized the Integral Basis to time-ordered integral in higher dimension by adopting the upper limit function trick, which is an innovative technique of numerical integration but have not been formally introduced.

We have introduced the Partition-Extrapolation methods which are algorithms for handling the early time divergence from the electrical engineering discipline. Remarkably, they are highly efficient in suppressing the divergence in short time period and we have tested for their fast convergence speed through our computation. Such efforts in the realm of electrical engineering could be paid more attention in a trans-disciplinary manner.

To visualize the performance of different integration methods, we have applied them to an analytical toy integral in the Quadratic Potential Single Field Inflation model and a numerical toy integral from the Feature Potential Single Field Inflation model. Among four summation methods, we have chosen the Beta Regulator to be the representative because of its best performance among them. We have presented details of their integration result, computation time, and convergence speed. The Integral Basis generally requires the least computation time and the two Partition-Extrapolation methods are able to reach the highest precision. The Beta Regulator is simple to implement and it can achieve a balance between convergence speed and computation time.

\section*{Acknowledgments}
This work was supported in part by the National Key R\&D Program
of China (2021YFC2203100), the NSFC Excellent Young Scientist Scheme (Hong Kong and Macau) Grant No. 12022516, and CRF C6017-20GF, GRF 16303819 by the RGC of Hong Kong SAR.

\appendix
\section{Review of $i\epsilon$ prescription in QFT}
By using the approach of path integral, we use  $\ket{n}$, $\ket{n'}$ to denote the energy eigenstate and $E_{n}$, $E_{n'}$ to denote the eigenvalue of the Hamiltonian H in the free theory and full theory, respectively. The $\ket{0}$ denotes the vacuum state of the free theory and the $\ket{\Omega}$ denotes the interaction vacuum state. One can expand $\ket{0}$ by a complete set of $\ket{n}$ near $\tau_{0} \to -\infty$ so that 
\begin{align}
\label{vacuumstate}
    e^{-iH(\tau-\tau_{0})}\ket{0}& = \sum_{n} e^{-i E_{n}(\tau-\tau_{0})} \ket{n} \bra{n}\ket{0} \notag \\
    &= e^{-i E_{\Omega}(\tau-\tau_{0})} \ket{\Omega} \bra{\Omega}\ket{0}+ \sum_{n^\prime} e^{-i E_{n^\prime}(\tau-\tau_{0})} \ket{n^\prime} \bra{n^\prime}\ket{0}~.
 \end{align}
To extract the ground state we make the replacement of time by adding an infinitesimal imaginary part to it
\begin{equation}
\label{imaginaryrotation}
\tau \to \tau (1-i\epsilon)~.
\end{equation}
One can obtain that only the term $\ket{\Omega}$ remains in \eqref{vacuumstate} in the limit $\tau_{0}\to-\oo(1-i\e)\equiv-\oo^{-}$. The ground state can then be expressed as
\begin{equation}
\label{groundstate}
\hspace{-2mm}\lim_{\tau_{0}
\to-\oo^{-}}e^{-iH(\tau-\tau_{0})} \ket{\Omega} = \lim_{\tau_{0}
\to-\oo^{-}}\frac{e^{-iH(\tau-\tau_{0})}\ket{0}}{\bra{\Omega}\ket{0}}~.
\end{equation}
Similarly, for any time evolution operator in the interaction picture, it can be written as
\begin{equation}
\label{interactionoperator}
\setlength{\abovedisplayskip}{12pt}
F(\tau,-\oo^{-}) \ket{\Omega} = \frac{F(\tau,-\oo^{-})\ket{0}}{\bra{\Omega}\ket{0}}~,
\end{equation}
\noindent where
\begin{equation}
\label{timeorderinteraction}
F(\tau,-\oo^{-})  = T e^{-i \int_{-\oo}^{\tau} H_I (\tau^{\prime})d\tau^{\prime}}~.
\end{equation}
\noindent The expectation value in the in-in formalism of an operator $W(\tau)$ can be expressed as 

\begin{equation}
\label{ininexpectation}
 \<W(\tau)\> = \left\langle\Omega\middle|\left [Te^{i\int_{\tau_{0}}^{\tau}H_{I}(\tau^{\prime})d\tau^{\prime}}\right]^{\dagger} W(\tau) 
\left [Te^{i\int_{\tau_{0}}^{\tau}H_{I}(\tau^{\prime})d\tau^{\prime}}\right]\middle|\Omega\right\rangle~.
\end{equation}
with respect to the initial state and the perturbation series, $\ket{\Omega}$, the interaction vacuum, and also often referred as $\ket{in}$. Now combine (\ref{groundstate}) and (\ref{timeorderinteraction}) we can write the in-in expectation value as
\begin{equation}
\label{combinedininexpectation}
 \<W(\tau)\> = \frac{\left\<0\middle|{\left [Te^{i\int_{\tau_{0}}^{\tau}H_{I}(\tau^{\prime})d\tau^{\prime}}\right]}^{\dagger} W(\tau) \left [Te^{i\int_{\tau_{0}}^{\tau}H_{I}(\tau^{\prime})d\tau^{\prime}}\right]\middle|0 \right\>}{{|\bra{0}\ket{\Omega}|}^{2}}~.
\end{equation}

\section{The upper limit function trick for time-ordered factorizable integrals}
\label{ulftrick}
To illustrate the idea of the upper limit function trick, we consider a general numerical time-ordered factorizable double integral
\begin{align}
    \label{doubleintegral}
   I_{2D} =  \int_{\tau_i}^{\tau_f}\int_{\tau_i}^{\tau_1}f_{1}(\tau_{1})f_{2}(\tau_{2})d\tau_{2}d\tau_{1}~,
\end{align}
where $\tau_i$ is in the limit of early time, i.e., $\tau_i \to -\oo$ in the numerical computation. Traditionally, the computation of a two-dimension integral requires to partition the integration domain into small squares within which the integrand is evaluated. The number of evaluation times of the integrand is determined by the upper and lower limit of integration accordingly. In $I_{2D}$, the evaluation times equal to $\frac{1}{2}\kappa(\tau_f - \tau_e)^2$, where $\kappa$ is a constant. Now, to effectively reduce the evaluation times and hence improve the computation speed we define the function
\begin{equation}
\label{Gtau1}
    F_{2}(\tau_1) = \int_{\tau_i}^{\tau_1} d\tau_2 f_{2}(\tau_2)~,
\end{equation}
which is the function of $\tau_1$ according to the Fundamental Theorem of Calculus and can be obtained by solving the differential equation with boundary conditions
\begin{equation}
\label{BVP}
    \frac{dF_{2}}{d\tau} = f_{2}(\tau), F_{2}(\tau_i) = 0, \tau \in [\tau_i, \tau_f]~.
\end{equation}
In solving this boundary value problem, $f_{2}(\tau)$ is approximately evaluated for $\kappa (\tau_f - \tau_i)$ times. Substituting $F_{2}(\tau_1)$ into $I_{2D}$, we can rewrite the integral as
\begin{equation}
\label{reviseddoubleintegral}
    I_{2D} =  \int_{\tau_i}^{\tau_f} f_{1}(\tau_1)F_{2}(\tau_1) d\tau_1~,
\end{equation}
and the evaluation times for the integral becomes $\kappa (\tau_f - \tau_i)$. The total integrand evaluation times of this process is $ 2\kappa(\tau_f-\tau_i) $. So, one can reduce the times of evaluating the integrand in the double or multi-dimensional time-ordered integral to scale up only linearly with the length of the integration interval via the trick shown above.

This trick can be used to reduce the computation time of the Hölder and Cesàro summation methods, which involves computing multi-dimensional time-ordered integrals.

In addition, the idea of computing the upper limit function is also useful in directly generalizing the damping factor method, boundary regulator method and integral basis method to multi-dimensional time-ordered factorizable integrals. However, we cannot directly generalize summation-based and partition-extrapolation methods to higher dimension. Instead, we can adopt a hybrid approach which is explained in Appendix \ref{ulftrickwithode}.

\section{A hybrid approach to implement the upper limit function trick}
\label{ulftrickwithode}
In general, we can generalize any method to multi-dimensional integrals by treating each layer of integral as an one-integral and evaluate them for each value of the other variables. The upper limit function trick in Appendix \ref{ulftrick} can reduce the complexity of computing time-ordered integrals. However, in summation-based and partition-extrapolation methods, the upper limit function trick cannot be implemented directly. This is because the upper limit of the integrals $\tau_{0}$ appears many times in the multi-dimensional integrals in \eqref{Holdersum} and \eqref{Cesarosum}, while in the Riesz summation, the upper limit $\tau_{0}$ in \eqref{Rieszsum} is mixed with the integrand $ f(\tau) $.

Therefore, we adopt the hybrid approach which allows generalizing any methods to handle $ i\epsilon $ to multi-dimensional time-ordered integrals. We use one method (let's call it Method X) to compute the value of the upper limit function at one point and let this value be the boundary condition for the boundary value problem for which we can solve to obtain the upper limit function on the whole integration interval. To illustrate, we consider a time-ordered two-integral
\begin{align}
    \label{model2-integral}
	\int_{-\oo}^{\tau_{0}}d\tau_{1}f_{1}(\tau_{1})\int_{-\oo}^{\tau_{1}}d\tau_{2}f_{2}(\tau_{2})~,
\end{align}
then the generalization to higher dimensions will be straightforward. We define the upper limit function of $ f_{2}(\tau_{2}) $
\begin{align}
	F_{2}(\tau_{1})=\int_{-\oo}^{\tau_{1}}d\tau_{2}f_{2}(\tau_{2})~,
\end{align}
and therefore we can use some method X to calculate $ F_{2}(\tau_{1}^{\text{ref}}) $ for some $ \tau_{1}^{\text{ref}} $. We denote the result obtained to be $ V_{2}^{\text{ref}} $.

\subsection{The straightforward way}

We can compute the upper limit function at any value inside the interval of interest $ [\tau_{\text{early}},\tau_{0}] $ by solving the boundary value problem
\begin{align}
    \label{straightforwardbvp}
	\dfrac{dF_{2}}{d\tau_{1}}=f_{2}(\tau_{1}),\text{ }F_{2}(\tau_{1}^{\text{ref}})=V_{2}^{\text{ref}}~.
\end{align}
Then, we can substitute $ F_{2} $ into
\begin{align}
	\int_{-\oo}^{\tau_{0}}d\tau_{1}f_{1}(\tau_{1})F_{2}(\tau_{1})~,
\end{align}
and apply Method X one more time.

\subsection{Using the Levin's equation}

Given the fact that the integrand in cosmological correlation function are oscillatory before horizon crossing. Thus, we can write
${f_{2}(\tau_{2})=g_{2}(\tau_{2})e^{ik_{2}\tau_{2}}}$, where $ k $ is a real number and $ g_{2}(\tau_{2}) $ is a non-oscillatory function.	

We can define $ p(\tau_{2}) $ as the solution of the Levin's equation \cite{levinrulepaper}
\begin{align}
    \label{levinequation}
	p'(\tau_{2})+ik_{2}p(\tau_{2})=g_{2}(\tau_{2})~.
\end{align}
Then, we have
\begin{align}
	\int_{a}^{b}d\tau_{2}f_{2}(\tau_{2})=p(b)e^{ik_{2}b}-p(a)e^{ik_{2}a}~.
\end{align}
If we can choose suitable initial condition for the Levin's equation so that $ p(a)e^{ik_{2}a}=0 $ when $ a=-\infty(1-i\epsilon) $, then we get the desired upper limit function
\begin{align}
	F_{2}(\tau_{1})=p(\tau_{1})e^{ik_{2}\tau_{1}}~.
\end{align}
The solution of the Levin's equation has the form
\begin{align}
\label{levinsolution}
	p(\tau_{2})=ce^{-ik_{2}\tau_{2}}+\tilde{p}(\tau_{2})~,
\end{align}
where $ c $ is an arbitrary constant and $ \tilde{p}(\tau_{2}) $ is the specific solution. In \cite{levinrulepaper}, Levin proved that $ \tilde{p}(\tau_{2}) $ is non-oscillatory (or slowly oscillatory) if $ g_{2}(\tau_{2}) $ is non-oscillatory (slowly-oscillatory). Multiply \eqref{levinsolution} by $e^{ik_2\tau_2}$ we have
\begin{align}
    p(\tau_{2})e^{ik_{2}\tau_{2}}=c+\tilde{p}(\tau_{2})e^{ik_{2}\tau_{2}}~.
\end{align}
The second term vanishes as $ \tau_{2}\to-\infty(1-i\epsilon)\equiv-\infty^{-} $ and therefore we need to choose the boundary condition so that $ c=0 $. We also note that
\begin{align}
	F(\tau_{1}^{\text{ref}})=p(\tau_{1}^{\text{ref}})e^{ik_{2}\tau_{1}^{\text{ref}}}-\lim_{\tau_{2}\to-\infty^{-}}p(\tau_{2})e^{ik_{2}\tau_{2}}=c+\tilde{p}(\tau_{1}^{\text{ref}})e^{ik_{2}\tau_{1}^{\text{ref}}}-c=\tilde{p}(\tau_{1}^{\text{ref}})e^{ik_{2}\tau_{1}^{\text{ref}}},
\end{align}
and if we choose the initial condition to be
\begin{align}
	p(\tau_{1}^{\text{ref}})=V_{2}^{\text{ref}}e^{-ik_{2}\tau_{1}^{\text{ref}}},
\end{align}
we will have: $ p(\tau_{1}^{\text{ref}})=\tilde{p}(\tau_{1}^{\text{ref}}) $ and thus $ c=0 $. The advantage of using Levin's equation to compute the upper limit function is that the solution of the Levin's equation is non-oscillatory. Thus, the computation time may be shorter than using the straightforward way.

\subsection{Choosing the reference point}

It is obvious to see that choosing a later reference point $\tau_{1}^{\text{ref}}$ will take more time to calculate $F_{2}(\tau_{1}^{\text{ref}})$. However, the reference point should be chosen to be the latest possible value: $\tau_{1}^{\text{ref}}=\tau_{0}$. This is because, to obtain an accurate solution for the upper limit function when solving \eqref{straightforwardbvp} or \eqref{levinequation}, the boundary condition needs to be as accurate as possible. Since every method requires taking the integrals over a long enough interval to get a good convergence, we need $\tau_{1}^{\text{ref}}$ to be late enough so that the computation of $F_{2}(\tau_{1}^{\text{ref}})$ is convergent enough to reach the accuracy.

\section{Convergence of Hölder, Cesàro, Riesz sums for the case $ g(\tau)=\tau^{n} $ ($n\in\mathbb{N}$)}
\label{H-C-Rconvergence}
The Cesàro sum $ (C,\alpha) $ is equivalent to and compatible with the Hölder sum $ (H,\alpha) $ \cite{hardydivseries} and therefore the integral $ \int_{-\infty}^{\tau_{0}}f(\tau)d\tau $ can be summed by $ (C,\alpha) $ if and only if it can be summed by $ (H,\alpha) $. Moreover, if the integral $ \int_{-\infty}^{\tau_{0}}f(\tau)d\tau $ can be summed by both $ (C,\alpha) $ and $ (H,\alpha) $ , then
\begin{align}
     \underset{(C,\alpha)}{\int_{-\infty}^{\tau_{0}}}f(\tau)d\tau= \underset{(H,\alpha)}{\int_{-\infty}^{\tau_{0}}}f(\tau)d\tau~.
\end{align}
The Cesàro sum $ (C,\alpha) $ is also equivalent to and compatible with the Riesz sum $ (R,\tau,\alpha) $ because of the Fubini's Theorem. Therefore, if we can prove the Cesàro sum converges to the value obtained by applying the $ i\epsilon $ prescription, it will also hold for the Hölder sum and the Riesz sum. 

Now, we consider the Cesàro sum
\begin{align}
    \underset{(C,\alpha)}{\int_{-\oo}^{\tau_{0}}}\tau^{n}e^{i\omega\tau}d\tau&=\lim_{\tau\to-\infty}-\dfrac{\alpha!}{(\tau-\tau_{0})^{\alpha}}\int_{\tau_{0}}^{\tau}d\tau^{(\alpha)}\cdots\int_{\tau_{0}}^{\tau^{(2)}}d\tau^{(1)}\int_{\tau_{0}}^{\tau^{(1)}}d\tau^{(0)}(\tau^{(0)})^{n}e^{i\omega\tau^{(0)}}~.
\end{align}
Evaluating the first layer of integral from the anti-derivative, we get
\begin{align}
    \int_{\tau_{0}}^{\tau^{(1)}}d\tau^{(0)}(\tau^{(0)})^{n}e^{i\omega\tau^{(0)}}=P_{0}(\tau^{(1)})e^{i\omega\tau^{(1)}}-C_{0}~,
\end{align}
where $ P_{0} $ is some polynomial of degree $ n $ such that $ \frac{d}{d\tau}(P_{0}(\tau)e^{i\omega\tau})=\tau^{n}e^{i\omega\tau} $ and $ C_{0}\equiv P_{0}(\tau_{0})e^{i\omega\tau_{0}} $ is a constant. By letting $ \tau\to-\infty(1-i\epsilon) $, we see that
\begin{align}
    C_{0}=\int_{-\oo(1-i\epsilon)}^{\tau_{0}}\tau^{n}e^{i\omega\tau}d\tau~,
\end{align}
which means that $ C_{0} $ is the value of the integral obtained by applying the $ i\epsilon $ prescription. Now, we substitute the result of the first layer into the second layer and perform the integration using the anti-derivative. The result is
\begin{align}
    \int_{\tau_{0}}^{\tau^{(2)}}d\tau^{(1)}\int_{\tau_{0}}^{\tau^{(1)}}d\tau^{(0)}(\tau^{(0)})^{n}e^{i\omega\tau^{(0)}}=P_{1}(\tau^{(2)})e^{i\omega\tau^{(2)}}-C_{1}-C_{0}(\tau^{(2)}-\tau_{0})~,
\end{align}
where $ P_{1} $ is some polynomial of degree $ n $ such that $ \frac{d}{d\tau}(P_{1}(\tau)e^{i\omega\tau})=P_{0}(\tau)e^{i\omega\tau} $ and $ C_{1}\equiv P_{1}(\tau_{0})e^{i\omega\tau_{0}} $ is a constant. Inductively, we have
\begin{align}
    \int_{\tau_{0}}^{\tau}d\tau^{(\alpha)}\cdots\int_{\tau_{0}}^{\tau^{(2)}}d\tau^{(1)}\int_{\tau_{0}}^{\tau^{(1)}}d\tau^{(0)}(\tau^{(0)})^{n}e^{i\omega\tau^{(0)}}=P_{\alpha}(\tau)e^{i\omega\tau}-\sum_{j=0}^{\alpha}C_{\alpha-j}\dfrac{(\tau-\tau_{0})^j}{j!}~,
\end{align}
where $ P(\alpha) $ is some polynomial of degree $ n $. Therefore,
\begin{align}
    \label{Cesarolimit}
    \underset{(C,\alpha)}{\int_{-\oo}^{\tau_{0}}}\tau^{n}e^{i\omega\tau}d\tau&=\lim_{\tau\to-\oo}\left(-\dfrac{\alpha!P_{\alpha}(\tau)e^{i\omega\tau}}{(\tau-\tau_{0})^{\alpha}}+C_{0}+\sum_{j=1}^{\alpha}C_{j}\dfrac{\alpha!}{(\alpha-j)!(\tau-\tau_{0})^{j}}\right)~,
\end{align}
where we see that the second term converges to $ C_{0} $ and the third term converges to $ 0 $. Since $ P_{\alpha}(\tau) $ is a polymomial of degree $ n $, the first term is convergent if and only if $ \alpha>n $. Given that this condition is satisfied, the first term would converge to $ 0 $. Thus
\begin{align}
     \underset{(C,\alpha)}{\int_{-\oo}^{\tau_{0}}}\tau^{n}e^{i\omega\tau}d\tau&=C_{0}=\int_{-\oo(1-i\epsilon)}^{\tau_{0}}\tau^{n}e^{i\omega\tau}d\tau~,
\end{align}
as desired.

\section{Comparison between summation methods}
\label{comparisonofsummationmethods}
In this section, we will prove, both analytically and numerically, that the Beta Regulator has the best performance among the summation methods.

\subsection{Convergence speed of Cesàro/Riesz summation method}
\label{Rconvspeed}
The result in Appendix \ref{H-C-Rconvergence} also helps us estimate the convergence speed of the Cesàro sum (and also the Riesz sum, since they are identical). Indeed, we see that the first term in \eqref{Cesarolimit} is $ O(\tau^{n-\alpha}) $ and the third term in \eqref{Cesarolimit} is $ O(\tau^{-1}) $. Since $ n-\alpha<-1 $, then the overall convergence speed is $ O(\tau^{-1}) $ for all $ \alpha>n $. Therefore, we cannot improve the convergence speed by increasing $ \alpha $. However, increasing $ \alpha $ can suppress the oscillatory term (the first term) in \eqref{Cesarolimit} more quickly.

A way to improve the convergence speed is by taking a weighted mean of Cesàro/Riesz sums. We note that, if an integral can be summed by $ (C,\alpha) $, it can be summed by $ (C,\alpha') $ as well, with the $ (C,\alpha') $ sum being equal to the $ (C,\alpha) $ sum, for all $ \alpha'>\alpha $ \cite{hardydivseries}. Therefore, any weighted mean of $ (C,\alpha) $ and $ (C,\alpha') $ sums will also be equal to the $ (C,\alpha) $ sum. 

From the proof in Appendix \ref{H-C-Rconvergence}, we notice that the coefficients $ C_{j} $ depends only on $ f $ and $ j $, not on $ \alpha $. The term $ \tau^{-1} $ in \eqref{Cesarolimit} is
\begin{align}
    \frac{\alpha}{\tau-\tau_{0}}C_{1}~.
\end{align}
We consider the weighted mean
\begin{align}
    \dfrac{1}{1-\frac{\alpha}{\alpha'}}\lim_{\tau\to-\infty}\int_{\tau}^{\tau_{0}}\left[\left(1-\dfrac{\tau'-\tau_{0}}{\tau-\tau_{0}}\right)^{\alpha}-\frac{\alpha}{\alpha'}\left(1-\dfrac{\tau'-\tau_{0}}{\tau-\tau_{0}}\right)^{\alpha'}\right]f(\tau')d\tau'~, 
\end{align}
where $ \alpha'>\alpha $, and the linear combination of the $ \tau^{-1} $ terms
\begin{align}
    \frac{\alpha}{\tau-\tau_{0}}C_{1}-\frac{\alpha}{\alpha'}\frac{\alpha'}{\tau-\tau_{0}}C_{1}=0~.
\end{align}
The $ \tau^{-1} $ term cancels out. Therefore, if we choose $ \alpha>n+1 $, we can achieve convergence speed $ O(\tau^{-2}) $. Similarly, we can take other weighted means to cancel more terms in $ \eqref{Cesarolimit} $ and further improve the convergence speed. One of such weighted means is \eqref{generalmodifiedRieszsum}, whose fast convergence will be proved here. Firstly, we note that the oscillating term in \eqref{Cesarolimit} will converge at rate $ O(\tau^{-q}) $ when $ \alpha>n+q-1 $. Since $ \alpha>q-1 $ can be guaranteed when $ n\geq0 $, so that at least $ q-1 $ terms are present in the sum in \eqref{Cesarolimit}. Then, we prove the identity: For positive integers and $ j\geq1 $ and $ \alpha>j $, we have
\begin{align}
    \label{coreidentitymodriesz}
    \dfrac{\alpha!}{(\alpha-j)!}-\dfrac{\alpha}{\alpha+1}\cdot\dfrac{(\alpha+1)!}{(\alpha+1-j)!}=(j-1)\dfrac{\alpha!}{[\alpha-(j-1)]!}~,
\end{align}
and indeed, we see that
\begin{align}
    LHS=\dfrac{\alpha!}{(\alpha-j)!}\left(1-\dfrac{\alpha}{\alpha-j+1}\right)=\dfrac{\alpha!}{(\alpha-j)!}\cdot\dfrac{j-1}{\alpha-j+1}=RHS~.
\end{align}
Next, we apply identity \eqref{coreidentitymodriesz} inductively to construct an explicit linear combination of Riesz sum which cancels the first $ q-1 $ terms in the sum \eqref{Cesarolimit}. For simplicity, here we denote the Cesàro/Riesz sum by $ (C^{(1)},\alpha) $. We define the linear combination
\begin{align}
    (C^{(2)},\alpha)\equiv(C^{(1)},\alpha)-\dfrac{\alpha}{\alpha+1}(C^{(1)},\alpha+1)~.
\end{align}
We substitute the RHS with \eqref{Cesarolimit} and use \eqref{coreidentitymodriesz} to simplify the sum that
\begin{align}
    (C^{(2)},\alpha)=\lim_{\tau\to-\infty}\left(O((\tau-\tau_{0})^{-q})\times e^{i\omega\tau}+\dfrac{C_{0}}{\alpha+1}+\sum_{j=2}^{\alpha}\dfrac{C_{j}(j-1)}{(\tau-\tau_{0})^{j}}\dfrac{\alpha!}{[\alpha-(j-1)]!}\right)~.
\end{align}
We can absorb $ (j-1) $ into $ C_{j} $ to form a new coefficient $ C'_{j} $ which is still independent of $ \alpha $ and only depends on $ f $ and $ j $. Then, we shift the sum variable to obtain
\begin{align}
    (C^{(2)},\alpha)=\lim_{\tau\to-\infty}\left(O((\tau-\tau_{0})^{-q})\times e^{i\omega\tau}+\dfrac{C_{0}}{\alpha+1}+\sum_{j=1}^{\alpha-1}\dfrac{C'_{j+1}}{(\tau-\tau_{0})^{j+1}}\dfrac{\alpha!}{(\alpha-j)!}\right)~.
\end{align}
We now see that the pattern of coefficients in the sum in the RHS is exactly the same as the pattern in \eqref{Cesarolimit}. As a result, we can keep canceling more and more terms by defining inductively
\begin{align}
    (C^{(m+1)},\alpha)\equiv(C^{(m)},\alpha)-\dfrac{\alpha}{\alpha+1}(C^{(m)},\alpha+1)~.
\end{align}
Eventually, only $ O((\tau-\tau_{0})^{-q}) $ terms and the constant term $ C_{0}\times\text{sum of weights} $ are left in $ (C^{(q)},\alpha) $. Next, we prove that
\begin{align}
    \label{weightsofmodriesz}
    (C^{(m)},\alpha)=\sum_{i=0}^{m-1}(-1)^{i}\binom{m-1}{i}\dfrac{\alpha}{\alpha+i}(C^{(0)},\alpha+i)~,
\end{align}
which confirms the weights chosen in \eqref{generalmodifiedRieszsum}. We easily see that the identity holds for the base case $ m=1 $. Assume this holds up until $ m=k $. Then
\begin{align}
    (C^{(k+1)},\alpha)=&(C^{(k)},\alpha)-\dfrac{\alpha}{\alpha+1}(C^{(k)},\alpha+1)\notag\\
    =&\sum_{i=0}^{m-1}(-1)^{i}\binom{m-1}{i}\dfrac{\alpha}{\alpha+i}(C^{(0)},\alpha+i)\notag\\
    &-\dfrac{\alpha}{\alpha+1}\sum_{i=0}^{m-1}(-1)^{i}\binom{m-1}{i}\dfrac{\alpha+1}{\alpha+1+i}(C^{(0)},\alpha+1+i)\notag\\
    =&(C^{(0)},\alpha)+(-1)^{m}\dfrac{\alpha}{\alpha+m}(C^{(0)},\alpha+m-1)\notag\\
    &+\sum_{i=1}^{m-1}(-1)^{i}\left[\binom{m-1}{i}+\binom{m-1}{i-1}\right]\dfrac{\alpha}{\alpha+i}(C^{(0)},\alpha+i)\notag\\
    =&\sum_{i=0}^{m}(-1)^{i}\binom{m}{i}\dfrac{\alpha}{\alpha+i}(C^{(0)},\alpha+i)~.
\end{align}
In the fifth line, we shifted the summation variable of the sum in the third line to group terms with the same Cesàro sums together. In the last line, we use the Pascal's rule. This proves \eqref{weightsofmodriesz} by induction. Finally, in order to perform a weighted mean, the sum of the weights must be non-zero. We start by proving
\begin{align}
    \sum_{i=0}^{q-1}(-1)^{i}\binom{q-1}{i}\frac{\alpha}{\alpha+i}x^{\alpha+i}=\alpha B_{x}(\alpha,q)~,
\end{align}
where $ B_{x}(\alpha,q) $ is the incomplete beta function. Taking the derivatives of the LHS, we have
\begin{align}
    \dfrac{\partial}{\partial x}(LHS)=\alpha x^{\alpha-1}\sum_{i=0}^{q-1}(-1)^{i}\binom{q-1}{i}x^{i}=\alpha x^{\alpha-1}(1-x)^{q-1}=\dfrac{\partial}{\partial x}(\alpha B_{x}(\alpha,q))~.
\end{align}
The last equality follows from the definition of the incomplete beta function. Both the LHS and the RHS are equal to $ 0 $ at $ x=0 $. Thus, the identity is proved. Substituting $ x=1 $, we have
\begin{align}
    \sum_{i=0}^{q-1}(-1)^{i}\binom{q-1}{i}\frac{\alpha}{\alpha+i}=\alpha B(\alpha,q)=\alpha\dfrac{(\alpha-1)!(q-1)!}{(\alpha+q-1)!}=\left[\binom{\alpha+q-1}{\alpha}\right]^{-1}\neq0~,
\end{align}
where $ B(\alpha,q)\equiv B_{1}(\alpha,q) $ is the beta function. From this proof, it is clear that the identity \eqref{rieszfactorandbeta} holds.

\subsection{Convergence speed of Hölder summation method}
\label{Hconvspeed}

We still consider the case $ g(\tau)=\tau^{n} $ and assume that $ \alpha>n $, so that the Hölder sum converges. By performing the integrals sequentially like the Cesàro sum where $ g(\tau)=\tau^{n} $, we arrive at
\begin{align}
    \label{Holderlimit}
    &\dfrac{1}{\tau-\tau_{0}}\int_{\tau_{0}-\gamma}^{\tau}d\tau^{(\alpha)}\dfrac{1}{\tau^{(\alpha)}-\tau_{0}}\cdots\int_{\tau_{0}-\gamma}^{\tau^{(3)}}d\tau^{(2)}\dfrac{1}{\tau^{(2)}-\tau_{0}}\int_{\tau_{0}}^{\tau^{(2)}}d\tau^{(1)}\int_{\tau_{0}}^{\tau^{(1)}}d\tau^{(0)}f(\tau^{(0)})\notag\\
    =&F(\tau)+C_{0}+\dfrac{C_{1}}{\tau-\tau_{0}}+\sum_{j=2}^{\alpha}\dfrac{C_{j}'}{(j-1)!}\dfrac{[\ln(\tau_{0}-\tau)]^{j-1}}{\tau-\tau_{0}}~,
\end{align}
where $ C_{0},C_{1} $ are the same as $ C_{0},C_{1} $ in \eqref{Cesarolimit}, $ C_{j}' $ are some constants and $ F(\tau) $ is a function satisfying $ F(\tau)\sim O((\tau-\tau_{0})^{n-\alpha}e^{i\omega\tau}) $. Note that we have used the fact that the regularization constant $ \gamma $ is very small to simplify the expression. However, we should also note that the value of $ \gamma $ affects the value of $ C_{j}' $ ($j=1,...,\alpha$). As $ \tau\to-\infty $, the convergence speed is determined by the last term, which is the slowest convergent term with convergence rate $ O\left([\ln(\tau_{0}-\tau)]^{\alpha-1}/(\tau-\tau_{0})\right) $. Therefore, increasing $ \alpha $ will actually lower the convergence speed of the Hölder sum.

\subsection{Numerical example}
\begin{figure}[H]
    \centering
   \includegraphics[width=\textwidth]{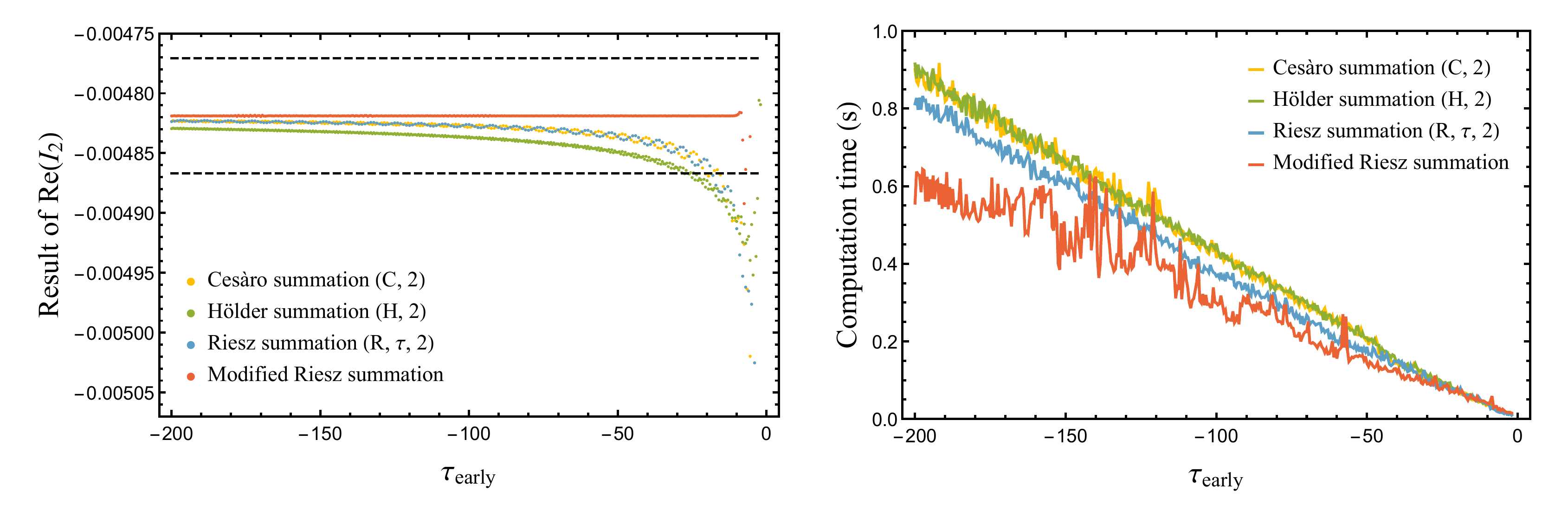}
  
    \caption{The dependence of the integration value (left) and the computation time (right) via four summation methods on the early time cut-off $\tau_{\text{early}}$. The dashed line represents the 1\% error boundary.}
    \label{fig:summationcurves}
\end{figure}

To visualize the performance of convergence through different summation methods, we apply them to the computation of the numerical toy integral \eqref{FMtoyintegral}. For this integral, since $ n=1 $, we choose $ \alpha=2 $ for the Hölder and Cesàro/Riesz sum. Among the Beta Regulator methods, we choose $ \alpha=20 $ and $ q=16 $ as the representative in this demonstration. The time-ordered integrals in the Hölder and Cesàro sums are implemented by using the upper limit function trick in \ref{ulftrick} to reduce the computation time. The dependence of the integration result on the cutoff value through four summation methods are summarized in the left graph in Fig.~\ref{fig:summationcurves}. The corresponding computation time at $\tau_{early}$ is shown in the right graph in Fig.~\ref{fig:summationcurves}.

This example confirms our previous expectations: all methods converge to the same limit; the Cesàro sum and Riesz sum are identical; the Beta Regulator converges faster than the Cesàro/Riesz sum, which in turn converges faster than the Hölder sum; the computation time of the Riesz sum is slightly lower than the Cesàro sum since it only involves one-dimensional integrals. Since the Beta Regulator achieves the best convergence speed, with no significant difference in computation time from the other summation methods, we choose this method to be the representative of summation methods to compare with other methods in Section \ref{comparison}.

\section{Convergence of the damping factor method}
\label{DFconv}

\begin{figure}[H]
    \centering
   \includegraphics[width=0.75\textwidth]{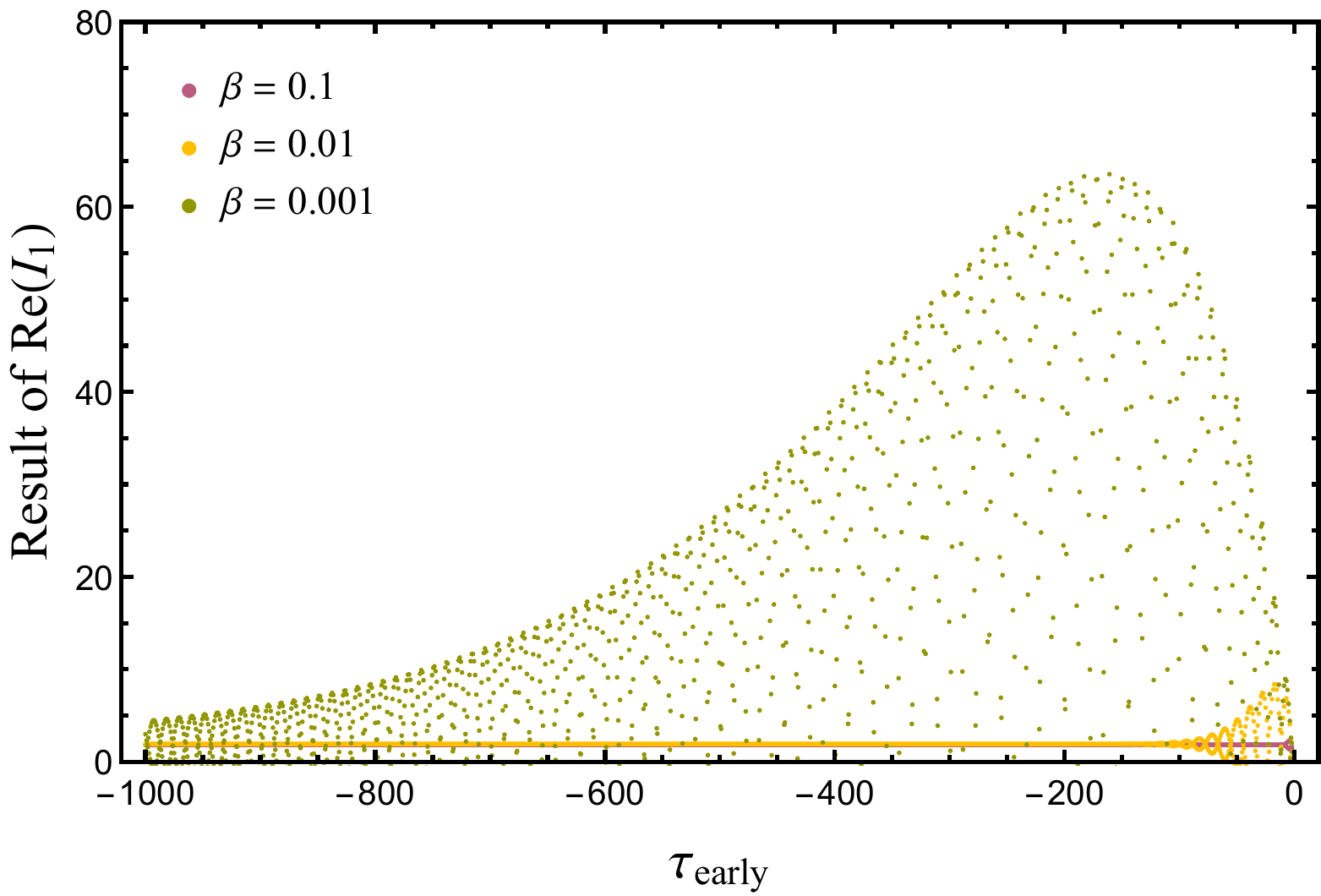}
  
    \caption{The dependence of the integration value via the damping factor methods on the early time cut-off $\tau_{\text{early}}$ at $\beta = 0.1, 0.01, 0.001$. }
    \label{fig:dampingfactor}
\end{figure}

\begin{figure}[htp]
    \centering
   \includegraphics[width=0.75\textwidth]{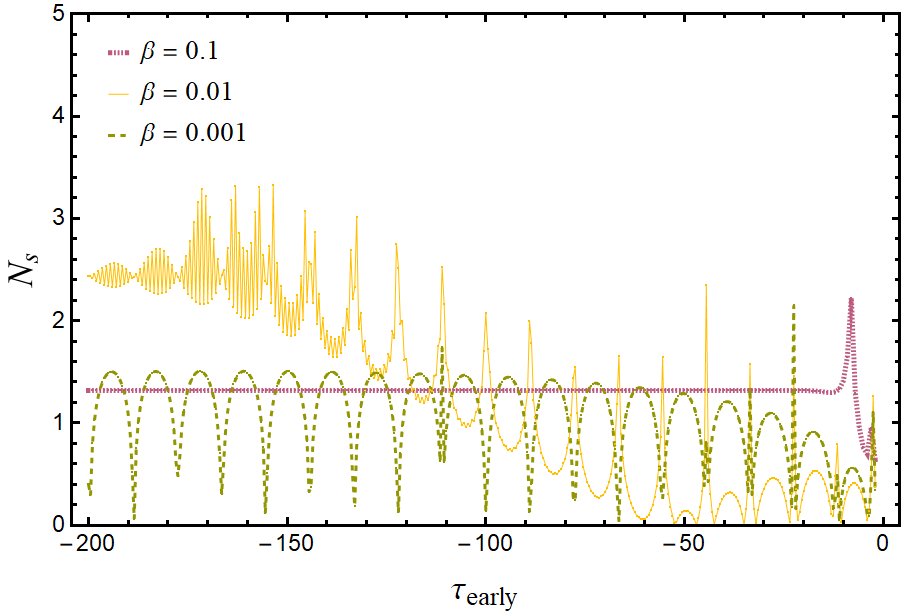}
  
    \caption{Comparison of the convergence ability of the damping factor method at $\beta = 0.1, 0.01, 0.001$.} 
    \label{fig:dampingfactordigits}
\end{figure}

In \ref{dampingfactor}, the damping factor method suggests that by adding a small $\beta$ into the integrand, one can suppress the early-time divergence by the $\beta$ appeared in exponential factor in \eqref{dampingintegrand}. The method holds only for small $\beta$ as the corresponding degree of rotation of $\tau$ into imaginary plane by $\tau \to \tau (1-i\beta)$ should remain minute to maintain the integration accurate. However, small value of $\beta$ may be problematic on the other hand as $e^{\beta\omega(\tau-\tau_0)}$ is not capable of effectively suppressing the divergence at early $\tau$. In general, the damping factor method returns accurate result when these conditions are satisfied \cite{chen2007large}
\begin{align}
    \beta\ll1\text{ and }|\beta\omega(\tau_{\text{early}}-\tau_{0})|\gg1~.
\end{align}
These conditions make the convergence of damping factor method extremely slow, although asymptotically, it is an exponential convergence. To illustrate this trade-off effect, we use the damping factor method at three different $\beta$ values namely 0.1, 0.01, 0.001 to compute the \eqref{integrandI} and the corresponding convergence of result (real) is shown in Fig.~\ref{fig:dampingfactor}. By plotting the $N_s$ with respect to $\tau_{early}$, the related convergence speed and the accuracy of the result is shown in Fig.~\ref{fig:dampingfactordigits}. We see that at small $\beta$, the integration result converges slowly while for large $\beta$, the accuracy of the result may be affected.

\section{More technical details of implementation}
\label{technicaldetails}

In this section, we provide some technical details encountered when implementing different methods. We also summarize notable characteristics of different methods.

In the boundary regulator method, we need to separate the non-oscillatory part $ g(\tau) $ of the integrand $f(\tau)$. If we know the frequency of oscillation $\omega$, we can readily obtain $ g(\tau)=f(\tau)e^{-i\omega\tau} $. Therefore, we need a prior knowledge of frequency. In the case of computing the cosmological correlation functions, the oscillation in the integrand comes from the mode functions, and we know the frequency of each Fourier mode of the fields.

In the Hölder summation method, since the integrand in each layer starting from the $\tau^{(2)}$-layer diverges at $\tau_{0}$, we can add a very small regularization term $\gamma=1\times10^{-15}$ into the lower limit of each integral except the first two:
\begin{align}
    \label{Holdersumregulated}
    \dfrac{1}{\tau_{\text{early}}-\tau_{0}}\int_{\tau_{0}-\gamma}^{\tau_{\text{early}}}d\tau^{(\alpha)}\dfrac{1}{\tau^{(\alpha)}-\tau_{0}}\cdots\int_{\tau_{0}-\gamma}^{\tau^{(3)}}d\tau^{(2)}\dfrac{1}{\tau^{(2)}-\tau_{0}}\int_{\tau_{0}}^{\tau^{(2)}}d\tau^{(1)}\int_{\tau_{0}}^{\tau^{(1)}}d\tau^{(0)}f(\tau^{(0)})~.
\end{align}

In the integral basis method, we also need to separate the non-oscillatory part $ g(\tau) $ of the integrand, so that we can fit it with a polynomial through linear regression. Consequently,the prior knowledge of frequency of the integrand is also required. Once we separate $g(\tau)$, it is able to determine the degree of $ g(\tau) $ ($n$ in \eqref{earlytimeapp}) by linearly fitting $\ln|g(\tau)| $ with $ \ln|\tau| $ in the early time region and determine $ n $ from the slope of the line fitted. After knowing $n$, we can fit $g(\tau)$ with the polynomial of degree $n$: $\sum_{m=0}^{n}c_{m}\tau^{m}$ to obtain the polynomial approximation.

Here is a summary:
\begin{table}[H]
\centering
    \begin{tabular}{|m{4.0cm}|>{\centering}m{2.8cm}|>{\centering}m{2.7cm}|>{\centering\arraybackslash}m{3.7cm}|}
    \hline
    \vspace{1mm}~~~~~~~~~~Method\vspace{1mm} &\vspace{1mm} Prior knowledge\vspace{1mm}&\vspace{1mm} Hyperparameter tuning\vspace{1mm} & \vspace{1mm}Convergence speed\vspace{1mm}\\
    \hline
   \vspace{1mm} Damping factor\vspace{1mm} &\vspace{1mm} No\vspace{1mm} &\vspace{1mm} $\beta$ \vspace{1mm}&\vspace{1mm} $\mathcal{O}(e^{\beta\omega\tau})$ (very slow)\vspace{1mm}\\
    \hline
    \vspace{1mm}Boundary regulator \vspace{1mm}&\vspace{1mm} Frequency\vspace{1mm} & \vspace{1mm}No\vspace{1mm} & \vspace{1mm}Power law with arbitrary power: $ \mathcal{O}((\tau-\tau_{0})^{p-m}) $\vspace{1mm}\\
    \hline
  \vspace{1mm} Beta regulator\vspace{1mm} & \vspace{1mm}No\vspace{1mm} & \vspace{1mm}No\vspace{1mm} &\vspace{1mm} Power law with arbitrary power\vspace{1mm}\\
    \hline
    \vspace{1mm}
    Integral basis\vspace{1mm} & \vspace{1mm}Frequency\vspace{1mm} & \vspace{1mm}No\vspace{1mm}& \vspace{1mm}$\mathcal{O}((\tau-\tau_{0})^{-1})$\vspace{1mm}\\
    \hline
   \vspace{1mm} Partition-extrapolation\vspace{1mm} & \vspace{1mm}Frequency\vspace{1mm}& \vspace{1mm}No\vspace{1mm} & \vspace{1mm}Fast\footnote{} \vspace{1mm}\\
    \hline
\end{tabular}
\caption{Summary of various characteristics corresponding to different methods.}
\end{table}
\footnotetext{The partition-extrapolation methods converge by annihilating one term from \eqref{remainderform} for each order of transformation \cite{pemethodsreview2}. The conventional sense of convergence speed based on asymptotic behavior is thus not applicable here.}

\pagebreak

\bibliographystyle{unsrt}

\begin{thebibliography}{10}

\bibitem{schwinger1960special}
Julian Schwinger.
\newblock The special canonical group.
\newblock {\em Proceedings of the national academy of sciences of the United
  States of America}, 46(10):1401, 1960.

\bibitem{bakshi1963expectation}
Pradip~M Bakshi and Kalyana~T Mahanthappa.
\newblock Expectation value formalism in quantum field theory. i.
\newblock {\em Journal of Mathematical Physics}, 4(1):1--11, 1963.

\bibitem{BakshiPradipM1963EVFi}
Pradip~M Bakshi and Kalyana~T Mahanthappa.
\newblock Expectation value formalism in quantum field theory. ii.
\newblock {\em Journal of mathematical physics}, 4(1):12--16, 1963.

\bibitem{Weinberg:2005vy}
Steven Weinberg.
\newblock {Quantum contributions to cosmological correlations}.
\newblock {\em Phys. Rev. D}, 72:043514, 2005.

\bibitem{Chen:2010xka}
Xingang Chen.
\newblock {Primordial Non-Gaussianities from Inflation Models}.
\newblock {\em Adv. Astron.}, 2010:638979, 2010.

\bibitem{Wang:2013zva}
Yi~Wang.
\newblock {Inflation, Cosmic Perturbations and Non-Gaussianities}.
\newblock {\em Commun. Theor. Phys.}, 62:109--166, 2014.

\bibitem{zhongzhi2017}
Xingang Chen, Yi~Wang, and Zhong-Zhi Xianyu.
\newblock {Schwinger-Keldysh Diagrammatics for Primordial Perturbations}.
\newblock {\em JCAP}, 12:006, 2017.

\bibitem{chen2010quasi}
Xingang Chen and Yi~Wang.
\newblock Quasi-single field inflation and non-gaussianities.
\newblock {\em Journal of Cosmology and Astroparticle Physics}, 2010(04):027,
  2010.

\bibitem{chen2007large}
Xingang Chen, Richard Easther, and Eugene~A Lim.
\newblock Large non-gaussianities in single-field inflation.
\newblock {\em Journal of Cosmology and Astroparticle Physics}, 2007(06):023,
  2007.

\bibitem{arroja2011large}
Frederico Arroja, Antonio~Enea Romano, and Misao Sasaki.
\newblock Large and strong scale dependent bispectrum in single field inflation
  from a sharp feature in the mass.
\newblock {\em Physical Review D}, 84(12):123503, 2011.

\bibitem{adshead2013bispectrum}
Peter Adshead, Wayne Hu, and Vin{\'\i}cius Miranda.
\newblock Bispectrum in single-field inflation beyond slow-roll.
\newblock {\em Physical Review D}, 88(2):023507, 2013.

\bibitem{chen2008generation}
Xingang Chen, Richard Easther, and Eugene~A Lim.
\newblock Generation and characterization of large non-gaussianities in single
  field inflation.
\newblock {\em Journal of Cosmology and Astroparticle Physics}, 2008(04):010,
  2008.

\bibitem{junaid2015geometrical}
M~Junaid and D~Pogosyan.
\newblock Geometrical measures of non-gaussianity generated from single field
  inflationary models.
\newblock {\em Physical Review D}, 92(4):043505, 2015.

\bibitem{pemethodsreview}
K.A. Michalski.
\newblock Extrapolation methods for sommerfeld integral tails.
\newblock {\em IEEE Transactions on Antennas and Propagation},
  46(10):1405--1418, 1998.

\bibitem{pemethodsreview2}
Krzysztof~A. Michalski and Juan~R. Mosig.
\newblock Efficient computation of sommerfeld integral tails – methods and
  algorithms.
\newblock {\em Journal of Electromagnetic Waves and Applications},
  30(3):281--317, 2016.

\bibitem{hardydivseries}
G.~H. (Godfrey~Harold) Hardy.
\newblock {\em Divergent series}.
\newblock Chelsea Pub. Co., New York, N.Y., 2nd (texually unaltered) ed..
  edition, 1991.

\bibitem{titchmarsh}
E.~C. (Edward~Charles) Titchmarsh.
\newblock {\em Introduction to the theory of Fourier integrals}.
\newblock Chelsea Pub. Co., New York, N.Y., 3rd ed.. edition, 1986.

\bibitem{Weniger:1989rea}
Ernst~Joachim Weniger.
\newblock {Nonlinear sequence transformations for the acceleration of
  convergence and the summation of divergent series}.
\newblock {\em Comput. Phys. Rept.}, 10(5-6):189--371, 1989.

\bibitem{levintranspaper}
David Levin.
\newblock Development of non-linear transformations for improving convergence
  of sequences.
\newblock {\em International Journal of Computer Mathematics}, 3(1-4):371--388,
  1972.

\bibitem{smithfordd-trans}
David~A. Smith and William~F. Ford.
\newblock Acceleration of linear and logarithmic convergence.
\newblock {\em SIAM Journal on Numerical Analysis}, 16(2):223--240, 1979.

\bibitem{Sidioriginalpaper}
Avram Sidi.
\newblock An algorithm for a special case of a generalization of the richardson
  extrapolation process.
\newblock {\em Numerische Mathematik}, 38(3):299--307, 1982.

\bibitem{sidiintegral1}
Avram Sidi.
\newblock A user-friendly extrapolation method for oscillatory infinite
  integrals.
\newblock {\em Mathematics of Computation}, 51(183):249--266, 1988.

\bibitem{sidiintegral2}
Avram Sidi.
\newblock {A user-friendly extrapolation method for computing infinite range
  integrals of products of oscillatory functions}.
\newblock {\em IMA Journal of Numerical Analysis}, 32(2):602--631, 08 2011.

\bibitem{maldacena2003non}
Juan Maldacena.
\newblock Non-gaussian features of primordial fluctuations in single field
  inflationary models.
\newblock {\em Journal of High Energy Physics}, 2003(05):013, 2003.

\bibitem{mathematicaprecision}
Wolfram Research.
\newblock Numerical operations on functions - wolfram language documentation,
  2022.
\newblock [Online; accessed 29-January-2022].

\bibitem{Mukhanov:1985rz}
Viatcheslav~F. Mukhanov.
\newblock {Gravitational Instability of the Universe Filled with a Scalar
  Field}.
\newblock {\em JETP Lett.}, 41:493--496, 1985.

\bibitem{levinrulepaper}
David Levin.
\newblock Procedures for computing one- and two-dimensional integrals of
  functions with rapid irregular oscillations.
\newblock {\em Mathematics of computation}, 38(158):531--538, 1982.

\end{thebibliography}

\end{document}